\documentclass[preprint,12pt]{elsarticle}

\usepackage{amsmath,amssymb}
\usepackage{booktabs}
\usepackage{siunitx}
\usepackage{hyperref}
\usepackage{longtable}
\usepackage{subcaption}
\usepackage{graphicx}

\journal{Acta Astronautica}

\begin{document}

\begin{frontmatter}



\title{Secular dynamics and the lifetimes of lunar artificial satellites under natural force-driven orbital evolution}


\author[UniGe]{Edoardo Legnaro}
\author[UniPD]{Christos Efthymiopoulos}

\affiliation[UniGe]{organization={University of Genova. Department of Mathematics},
            addressline={Via Dodecaneso 35}, 
            city={Genova},
            postcode={16146}, 
            state={GE},
            country={Italy}}

\affiliation[UniPD]{organization={University of Padova. Department of Mathematics "Tullio Levi Civita"},
            addressline={Via Trieste 63}, 
            city={Padova},
            postcode={35131}, 
            state={PD},
            country={Italy}}

\begin{abstract}
    In this paper, we study the long-term (time scale of several years) orbital evolution of lunar satellites under the sole action of natural forces. In particular, we focus on \textit{secular resonances}, caused either by the influence of the multipole moments of the lunar potential and/or by the Earth's and Sun's third-body effect on the satellite's long-term orbital evolution. Our study is based on a simplified secular model obtained in `closed form' using the same methodology proposed in the recently published report on the semi-analytical propagator of lunar satellite orbits, SELENA \cite{efthymiopoulos2023selena}. Contrary to the case of artificial Earth satellites, in which many secular resonances compete in dynamical impact, we give numerical evidence that for lunar satellites only the $2g-$resonance ($\dot{\omega} = 0$) affects significantly the orbits at secular timescales. We interpret this as a consequence of the strong effect of lunar mascons. We show that the lifetime of lunar satellites is, in particular, nearly exclusively dictated by the $2g$ resonance. By deriving a simple analytic model, we propose a theoretical framework which allows for both qualitative and quantitative interpretation of the structures seen in numerically obtained lifetime maps. This involves explaining the main mechanisms driving \textit{eccentricity growth} in the orbits of lunar satellites. In fact, we argue that the re-entry process for lunar satellites is not necessarily a chaotic process (as is the case for Earth satellites), but rather due to a sequence of bifurcations leading to a dramatic variation in the structure of the separatrices in the $2g$ resonance's phase portrait, as we move from the lowest to the highest limit in inclination (at each altitude) where the $2g$ resonance is manifested.
\end{abstract}


\end{frontmatter}


\section{Introduction}
\label{sec: introduction}
The exploration of the lunar space and the deployment of satellites around the Moon represent critical steps in space exploration. These endeavors are emphasized in the Global Exploration Roadmap \cite{NASAISECG2023}, a document outlining the collaborative efforts of international space agencies to set milestones for expanding human presence from Earth's orbit to that of the Moon or Mars. 

From a theoretical standpoint, the study of artificial satellites' motion around non-spherical primaries has evolved from considering only the perturbation due to the body's oblateness (the $J_2$ term) (\cite{king1958effect}, \cite{allan1970critical}, \cite{jupp1987critical}, \cite{coffey1986critical}) to progressively including more complex gravitational models to accurately describe the satellites' dynamics. Many studies (\cite{knevzevic1998orbit}, \cite{folta2006lunar}, \cite{abad2009analytical}, \cite{lara2009preliminary}, \cite{nie2018lunar}) have focused on zonal harmonics. This is because the problem's Hamiltonian, when averaged over the mean motion of the satellite, reduces to a one-degree-of-freedom system (see \cite{saedeleer2005complete}), which allows for the identification of frozen orbits that maintain constant the argument of pericenter and eccentricity. These orbits, particularly near-polar frozen ones, are crucial for the planning and success of lunar mapping missions (\cite{singh2019mission}, \cite{wang2024transfers}).

However, the Moon's gravitational field is quite complex due to the presence of lunar mascons (\cite{muller1968mascons}, \cite{konopliv2001recent}, \cite{zuber2013gravity}). According to the recently published report on the semi-analytical propagator of lunar satellites SELENA (see \cite{efthymiopoulos2023selena}), the accumulation of several high-degree ($n>10$) harmonics of the multipole expansion of the lunar potential implies that a reasonable secular model, eliminating all important short-period effects, is hard to obtain at altitudes $\lesssim 100$ km above the Moon's surface. As a rough guide to comparative force estimates, from Figure 7 of \cite{efthymiopoulos2023selena}, we deduce that the force due to lunar potential harmonics at a multipole degree as high as $n=10$ becomes equal to or smaller than the force due to the Earth's tide on the satellite only at altitudes exceeding $\sim 500$ km.
Based on this information, we roughly distinguish three zones where the dynamics can be called essentially secular, i.e., where models averaged with respect to the satellite's mean anomaly represent fairly well the true dynamics:
\begin{itemize}
    \item The \emph{Low-Altitude zone} (altitudes between 100 - 500 km), where the dynamics is dominated by several important zonal and tesseral harmonics of degree $n\leq 10$, while the Earth's tide is negligible.
    \item The \emph{Middle-Altitude zone} (altitudes between 500 - 5000 km), where the forces produced on the satellite by some particular low-degree harmonics compete in size with the force due to the Earth.
    \item The \emph{High-Altitude zone} (beyond 5000 km and up to about one third of the size of the Moon's Hill sphere, $\sim 20000$ km), where the force due to the Earth is dominant.
    Our analysis in the present paper applies to all three zones above, but most results of practical interest refer to the middle-altitude zone. At any rate, the overall conclusion is that gravitational models substantially more comprehensive than $J_2+C_{22}$ are required for accurately recovering the long-term dynamics of satellite orbits in the Moon's environment.
\end{itemize}

In many problems within celestial mechanics, resonances resulting from a commensurability relationship between the fundamental orbital frequencies play a key role in the system's dynamics. In the case of Earth satellites, particularly in the Middle Earth orbit (MEO) zone, it is well known that several so-called 'secular' resonances, i.e., resonances between the frequencies of precession of the satellite's argument of perigee $g$, longitude of the ascending node $h$, and the Moon's node with respect to the ecliptic plane—lead to significant effects with many potential applications (e.g., in satellite end-of-life disposal):  the \textit{eccentricity growth} effect. In short, the separatrices of these resonances are such that orbits moving near them exhibit a growth in eccentricity driven solely by natural forces, most importantly, the lunisolar tidal forces. In the case of the so-called $2g+h$ and $2g$ resonances, this growth can eventually lead to an eccentricity value large enough that the orbit's perigee reaches atmospheric re-entry (\cite{chao2004long}, \cite{rossi2008resonant}, \cite{alessi2014effectiveness}, \cite{gkolias2016order}, \cite{alessi2016numerical}, \cite{armellin2018optimal}, \cite{skoulidou2019medium}). Additionally, the overlap of such resonances gives rise to large domains of chaotic motion in which diffusive phenomena occur (\cite{Daquin_Gkolias_Rosengren_2018}, \cite{gkolias2019chaotic}, \cite{legnaro2023semianalytical}). Analytical studies have provided insights into the mechanisms driving these phenomena for Earth satellites (\cite{celletti2014dynamics}, \cite{daquin2016dynamical}, \cite{celletti2017dynamical}, \cite{daquin2021deep}, \cite{Legnaro_Efthymiopoulos_2022}, and \cite{legnaro2023orbital}).

In the present paper, we explore the effects of secular resonances on the long-term dynamics of a lunar satellite. The emphasis is, again, on possible applications of such a study to understanding re-entry through the eccentricity growth mechanism. Here, re-entry simply means collision with the Moon's surface, that is, a perigee lunicentric distance smaller than the Moon's radius $r_L$. By numerical experiments (see below), one can easily see that this condition can be fulfilled even if the (constant in time, under secular dynamics) satellite's semi-major axis $a$ is given a value substantially larger than $r_L$. 
To this end, in our study, we first compute numerical \textit{lifetime maps}. These maps show in color scale the time up to which the satellite survives in orbit above the Moon's surface for a suitably chosen grid of initial conditions in element space. The numerical maps are obtained with a highly accurate force model (see section \ref{sec: model}), but we then attempt to understand their main features using a much simpler analytical secular model of the equations of motion. Notwithstanding its simplicity, we find that our simple analytical secular model is sufficient for most practical purposes to interpret the lifetime maps obtained by the far more complex numerical model. 
In fact, the analytical secular model allows to build a theory, based on the structure of the separatrices of the $2g$-resonance, which reproduces quite accurately the borders of the domains in element space separating initial conditions leading to re-entry from those which do not.

The structure of the paper is as follows: Section 2 describes the force model, equations of motion, as well as the numerical lifetime maps obtained with the above model. Section 3 describes the analytical secular model used to interpret theoretically the lifetime maps. Section 4 summarizes our main conclusions from the present study.  

\section{Force Model and numerical lifetime maps}
\label{sec: model}

\subsection{Force model}
\label{ssec: forcemodel}
We adopt as a reference frame the Principal Axis Lunar Reference Frame (PALRF). This is a solidal frame with its origin at the barycenter of the Moon, the $z$-axis coinciding with the Moon's mean axis of revolution, and the $x$-axis normal to the $z$-axis, pointing towards the Moon's meridian with the largest equatorial radius. Owing to the Moon-Earth synchronous spin-orbit resonance, the PALRF's $x$-axis practically points, also, towards the average lunicentric position of the Earth in the same frame.

The Hamiltonian describing the motion of a lunar satellite in the PALRF frame is
\begin{equation}
    H = \frac{1}{2} \vec{p}^{\;2} - \vec{\omega}_L \cdot (\; \vec{r}\times \vec{p} \;) + V(\vec{r}, t)
\end{equation}
where $\vec{r}(t)$ is the lunicentric PALRF radius vector of the satellite, $\vec{p}(t)$ is the velocity vector of the satellite in a fictitious rest frame whose axes instantaneously coincide with the axes of the PALRF at time $t$,  $\vec{\omega}_L(t)$ is the Moon's angular velocity vector and $V$ is the potential
\begin{equation}
    V(\vec{r}, t) =  V_L(\vec{r}) + V_E(\vec{r}, t) + V_S(\vec{r}, t)~~,
\end{equation}
where $V_L, V_E, V_S$ are respectively the potential of the Moon, the Earth and the Sun (in this study, we do not consider the influence of the solar radiation pressure). The time $t=0$ corresponds to the Julian day JD2000 at 12.00 Noon (UTC). From Hamilton's equations we obtain 
\begin{align}
    \label{vel}
    \dot{\vec{r}}(t) &= \vec{p}(t)-\vec{\omega}_L(t)\times\vec{r}(t), \\
    \ddot{\vec{r}}(t) &= -\nabla V(\vec{r},t)-\dot{\vec{\omega}}_L\times\vec{r}
    -2\vec\omega\times\dot{\vec{r}} - \vec{\omega}_L\times ( \;\vec{\omega}_L\times\vec{r} \;)~. \nonumber
\end{align}
The lunar potential can be written as 
\begin{equation}\label{potmoon}
    \small
    V_L(\, \vec{r}\,) = -{\mathcal{G}M_L\over r}\sum_{n=0}^\infty
    \left({r_L\over r}\right)^n\sum_{m=0}^n P_{nm}(\sin\phi)
    [C_{nm}\cos(m\lambda)+S_{nm}\sin(m\lambda)]
    \end{equation}
where $\mathcal{G}$ is Newton's gravity constant, $M_L$ is the Moon's mass and $r_L$ its radius. The angles $\phi$ and $\lambda$ are the satellite's longitude and latitude and $r$ is the lunicentric satellite's distance. 
The functions $P_{nm}$ are normalized Legendre polynomials of degree $n$ and order $m$, and $C_{nm}$, $S_{nm}$ are the zonal ($m=0$) and tesseral ($m\neq 0$) coefficients of the lunar gravity potential. We consider the numerical values of these parameters as provided by the GRAIL mission \cite{zuber2013gravity}. 

\noindent
The Earth's tidal potential is given by
\begin{equation}\label{potearth}
    V_E(\vec{r},t) = -\mathcal{G}M_E
    \left({1\over\sqrt{r^2+r_E^2 - 2\vec{r}\cdot\vec{r}_E}}-
    {\vec{r}\cdot\vec{r}_E\over r_E^3}\right)~~,
\end{equation}
    where $\vec{r}_E(t)$ is the lunicentric PALRF radius vector of the Earth. Multipole expansion yields:
\begin{equation}\label{potearthmpole}
    V_E(\vec{r},t) = 
     V_{E}^{P0}\left(\vec{r}_E(t)\right)
    +V_{E}^{P2}\left(\vec{r},\vec{r}_E(t)\right)
    +V_{E}^{P3}\left(\vec{r},\vec{r}_E(t)\right)
\end{equation}
with
\begin{align}
    V_{E}^{P0}\left(\vec{r}_E(t)\right) &=-{\mathcal{G}M_E \over r_E(t)}, \\
    V_{E}^{P2}\left(\vec{r},\vec{r}_E(t)\right) &= {\mathcal{G}M_E \over r_E(t)}\left({1\over 2}{r^2\over r_E^2(t)}-{3\over 2}{(\vec{r}\cdot\vec{r}_E(t))^2\over r_E^4(t)}\right), \\
    V_{E}^{P3}\left(\vec{r},\vec{r}_E(t)\right) &=
    {\mathcal{G}M_E \over r_E(t)}
    \left(
    {3\over 2}{r^2(\vec{r}\cdot\vec{r}_E(t))\over r_E^4(t)}
    -{5\over 2}{(\vec{r}\cdot\vec{r}_E(t))^3\over r_E^6(t)}
    \right).
\end{align}
The term $V_{E}^{P0}$ can be omitted from the Hamiltonian since it does not depend on the satellite's radius vector $\vec{r}$ and, therefore, does not contribute to the equations of motion.
The terms $V_{E}^{P2}$, $V_{E}^{P3}$ are hereafter referred to as the \emph{quadrupolar} and \emph{octopolar} Earth's tidal terms respectively. 
 
\noindent
The contribution due to the Sun is
\begin{equation}\label{potsun}
V_S(\vec{r},t) = -\mathcal{G}M_S
\left({1\over\sqrt{r^2+r_S^2(t) - 2\vec{r}\cdot\vec{r}_S(t)}}-
{\vec{r}\cdot\vec{r}_S(t)\over r_S^3(t)}\right)
\end{equation}
where $\vec{r}_S(t)$ is the lunicentric PALRF radius vector of the Sun. This can be expanded in multipoles analogously to the Earth's tidal potential. 

It is possible to estimate the relative strength in acceleration of these forces with respect to the keplerian one $a_{Kep}$ using the formulas proposed in \cite{efthymiopoulos2023selena}. 
Regarding the Moon's potential, we can estimate the acceleration $\Delta a(n)$ generated by the sum of all $n$-th degree harmonics through the formula
\begin{equation}
    \label{eq: acc_harmonics}
    \frac{\Delta a(n)}{a_{Kep}} \sim \sum_{m=0}^n {(n+1) r_L^n\over r^n}\left(C_{nm}^2+S_{nm}^2\right)^{1/2}~~.
\end{equation}
For perturbations due to the Earth’s and Sun’s tides ($\Delta a_E$, $\Delta a_S$), as well as non-inertial forces ($\Delta a_{NI}$) we have
\begin{equation} \label{eq: acc_perturbations}
    {\Delta a_E\over a_{Kep}}\sim {M_E r^3\over M_L a_E^3}, \quad 
    {\Delta a_S\over a_{Kep}}\sim {M_S r^3\over M_L a_S^3},\quad
    {\Delta a_{NI}\over a_{Kep}}\sim {n_L^2 r^3\over {\cal G} M_L} = {(M_E+M_L) r^3\over M_L a_E^3}~~.
\end{equation}

Notice that, as explained in \cite{efthymiopoulos2023selena}, the synchronous rotation of the Moon implies that the apparent forces due to the Moon’s rotation are of a similar size to the Earth’s tidal force at all possible lunicentric distances 
$r$ that can be reached by a satellite.

\begin{figure}
    \centering
    \includegraphics[width = \textwidth]{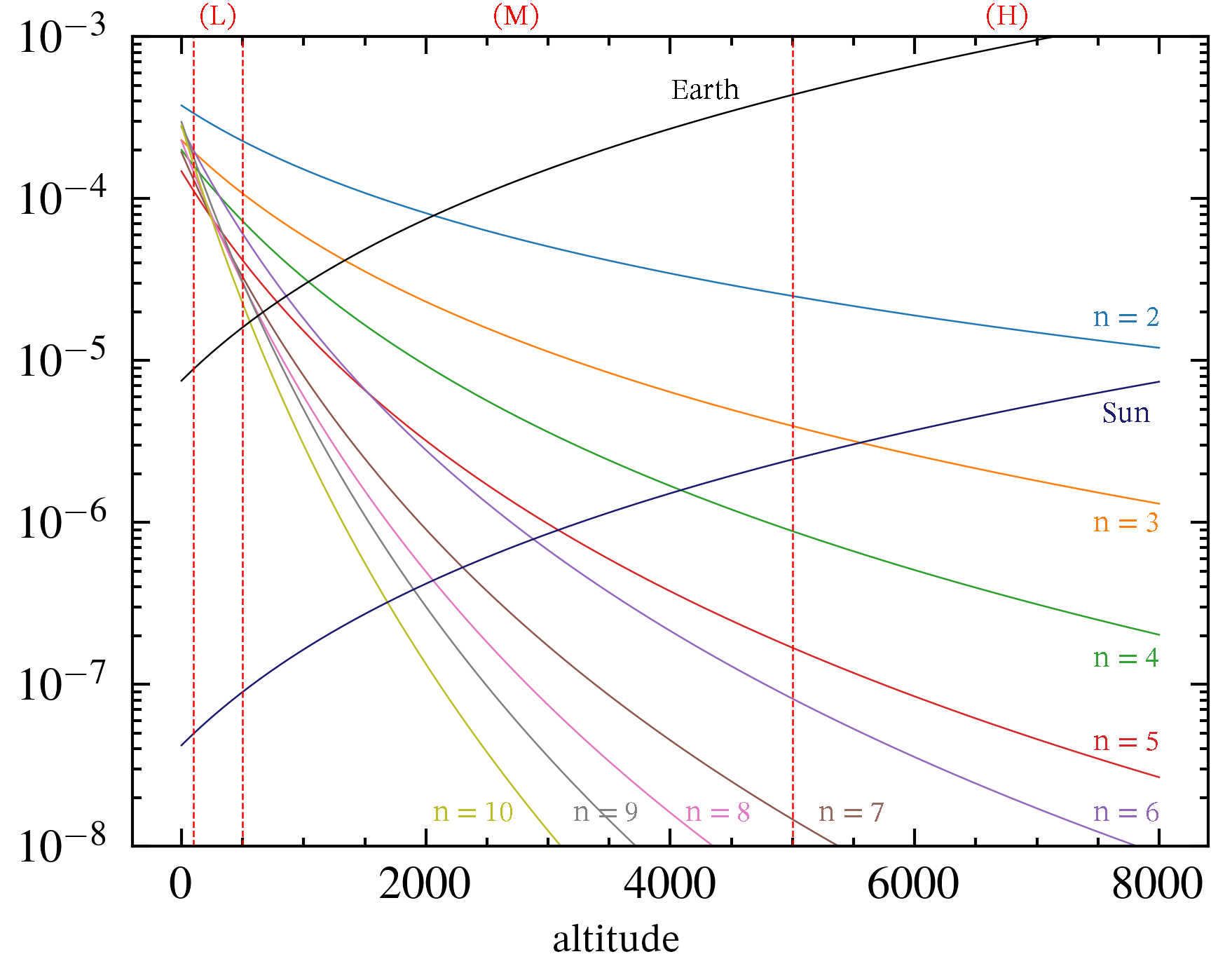}
    \caption{
        An estimate of the relative strength of the acceleration due to the lunar harmonics, the Earth and Sun with respect to the Keplerian acceleration as a function of the altitude (see Eq. \eqref{eq: acc_harmonics}, \eqref{eq: acc_perturbations}). The vertical red dashed lines (100 km, 500 km and 5000 km) indicate the limits of the various zones defined in the text according to altitude. 
        }
    \label{fig:moon_force}
\end{figure}
With these formulas, we get Figure \ref{fig:moon_force}, from which we can see that for low-altitude orbits many harmonics compete in size. Actually, it has been suggested that zonal terms up to $n=30$ still influence the secular behavior of orbits such as the low-altitude high-inclination ones (\cite{lara2020exploring}).
Also, notice that the Earth's tidal acceleration (and correspondingly apparent accelerations) becomes more prominent than second-degree lunar potential terms after an altitude of $2065$ km, while the Sun's contribution becomes important only at very high altitudes. 

Based on the information provided in Figure \ref{fig:moon_force}, we distinguish four zones in altitude, with somewhat arbitrary limits,  corresponding to different dynamical regimes regarding the effects and relative importance of the various force perturbations determining a satellite's long-term orbital dynamics.\\
\\
\noindent
{\it Zone of essentially non-secular dynamics} (altitude $a-r_L< 100$ km): this is the zone where lunar mascons dominate, implying that any secular force model, i.e., based on averaging of the equations of motion with respect to the satellite's mean anomaly,  would require canonical transformations including a large number of Lunar potential harmonics of degree higher than $n=10$. In practice, 
this means that, in osculating elements, the accumulation of the effects of all these harmonics would lead to a substantial (and possibly chaotic) long-term evolution of the satellite's semi-major axis, an effect not consistent with the definition of `secular' orbital behavior. \\
\\
\noindent
{\it Low Altitude Zone} of essentially secular dynamics (altitude $100 km\leq a-r_L< 500$ km): the low limit of this zone is somewhat arbitrary, but motivated by the work \cite{tzirtisecularlunar2014},  as well as by figures 14 and 15 of ref.\cite{efthymiopoulos2023selena}, which show that the semi-major axis variations become of order $<10^{-3}$, i.e, two orders of magnitude smaller than those required for re-entry. On the other hand, in this zone we have an accumulation of the contribution of many harmonics of degree $n\leq 10$, a fact leading to a total force perturbation of about two orders of magnitude larger than the one due to the Earth's tide.\\
\\
\noindent
{\it Middle Altitude Zone} of essentially secular dynamics (altitude $500 km\leq a-r_L\leq 5000$ km): near the lower limit of this zone the lunar multipoles produce a force perturbation superior by more than one order of magnitude to the Earth's tidal perturbation. However, this is reversed at the higher limit, where the problem becomes essentially one perturbed by the Earth plus few harmonics $n=2$ and $n=3$.\\
\\
\noindent
{\it High Altitude Zone} of essentially non-secular dynamics (altitude $a-r_L> 5000$ km): any perturbation other than the Earth's tidal becomes negligible. 

\subsection{Equations of motion}
\label{ssec: eqmo}
\noindent
Consider Delaunay variables
\begin{equation} \label{eq: delaunay}
    \small
    L = \sqrt{\mu_E a}, \quad G = L\sqrt{1-e^2}, \quad H= G \cos i, \quad \ell=M, \quad g=\omega, \quad h=\Omega,
\end{equation}
where $(\ell,g,h)=(M,\omega,\Omega)$ are the satellite's mean anomaly, argument of perilune and longitude of the nodes.
In view of the analysis done in the previous section, we compute secular equations of motion equivalent to those reported in \cite{efthymiopoulos2023selena} for the semi-analytical propagator of lunar satellites SELENA, but without the use of the normalizing transformation switching back and forth from mean to osculating elements. As explained in subsection 5.5 of \cite{efthymiopoulos2023selena}, a continuous transformation from mean to osculating elements along the orbital integration is not really necessary unless very high precision levels (better than one part in a million) are required, which is not the case in our present study. For orbits of relative error $\sim 10^{-6}$, in \cite{efthymiopoulos2023selena} it is shown that the crucial step is to make the canonical transformation from osculating to mean elements only as regards the initial datum. Then, propagating the orbit using the averaged equations of motion, and simply equating thereafter mean with osculating elements yields results equal up to six significant figures with the far more cumbersome 
approach of back-transforming from mean to osculating elements at every time step. This leads to a propagation model called `SELENA-mean' propagator in \cite{efthymiopoulos2023selena}. Here, we ignore even the initial transformation, which leads to orbits precise at about three significant figures, but with the advantage that one can simply reproduce the averaged equations of motion for all the terms included in the SELENA-mean model using just the basic formulas for first-order averaging of the Hamiltonian function in closed form. In addition, we ignore the SELENA terms due to the solar radiation pressure, which are orders of magnitude smaller than all other perturbations. This leads to the secular Hamiltonian model $\mathcal{H}_{SM}$, hereafter called the `SELENA-Model' (SM):
\begin{eqnarray}\label{eq: hamsm}
\mathcal{H}_{SM}(a,e,i,g,h)
&=&
-{\mathcal{G} M_L\over 2a}-\omega_{L,z}(t)H~~\nonumber\\
&-&
\bigg(
    \omega_{L,x}(t)\sin i\sin(h)
   -\omega_{L,y}(t)\sin i\cos(h)
\bigg)G~~\nonumber\\
&+&
{1\over 2\pi}\int_{0}^{2\pi} {a^2\over r^2\eta}V_L^{n\leq 10}df~\\
&+&
{1\over 2\pi}\int_{0}^{2\pi} (1-e\cos u)(V_E^{P2}+V_E^{P3}+V_S^{P2})du \nonumber
\end{eqnarray}
where $\eta=\sqrt{1-e^2}$ is the `eccentricity function' and $G$ and $H$ the Delaunay actions defined above. To obtain the integrals over the truncated multipole lunar potential in closed form, we use the formulas
\begin{eqnarray*}
&~&V_L^{n\leq 10}(a,e,i,f,g,h;t)= \\
&~&~~-{\mu_L\over r}\sum_{n=2}^{10}
    \left({r_L\over r}\right)^n\sum_{m=0}^n P_{nm}(\sin\phi)
    [C_{nm}\cos(m\lambda)+S_{nm}\sin(m\lambda)]
\end{eqnarray*}
with $\mu_L=\mathcal{G}M_L$ (the gravitational parameter of the Moon), setting
$$
\sin\phi=z/r
$$
$$
\cos(m\lambda)={1\over r^m}
\sum_{s=0}^{[m/2]} (-1)^s{m\choose 2s} x^{m-2s}y^{2s},~~
$$
$$\sin(m\lambda)={1\over r^m}
\sum_{s=0}^{[(m-1)/2]} (-1)^s{m\choose 2s+1} x^{m-2s-1}y^{2s+1}
$$
and then substituting $(x,y,z)$ by the equations (30) of \cite{efthymiopoulos2023selena}. Similarly, to obtain the integrals over the multipole Earth and Sun potential terms $V_E^{P2}$, $V_E^{P3}$ and $V_S^{P2}$, we use the formulas
\begin{equation}
    \begin{array}{rcl}
        r &=& a(1 - e \cos u) \\
        \vec{r} \cdot \vec{r}_E(t) &=& x~x_E(t) + y~y_E(t) + z ~z_E(t) \\
        \vec{r} \cdot \vec{r}_S(t) &=& x~x_S(t) + y~y_S(t) + z ~z_S(t)
    \end{array}
\end{equation}
with $(x,y,z)$ given by Eq. (37) of \cite{efthymiopoulos2023selena}. The Hamiltonian depends explicitly on time through the quantities 
\begin{equation}
    \begin{array}{rcl}
        \vec{\omega}_L(t)&=&(\omega_{L,x}(t), \omega_{L,y}(t),\omega_{L,z}(t)) \\
        \vec{r}_E(t)&=&(x_E(t),y_E(t),z_E(t)) ,\\
        \vec{r}_S(t)&=&(x_S(t),y_S(t),z_S(t)), 
    \end{array}
\end{equation}
which are respectively the Moon's angular velocity and the Earth and Sun vector radii in the PALRF system. The vectors $\vec{\omega}_L(t)$, $\vec{r}_E(t)$ and $\vec{r}_S(t)$  are coded using the equations (2), (10) and (13) of \cite{efthymiopoulos2023selena} with the accompanying tables of coefficients. 

While the Hamiltonian (\ref{eq: hamsm}) is formally expressed in Keplerian elements, in reality it is a function of the canonical Delaunay action-angle variables $(L, \ell)$, $(G, g)$, and $(H, h)$, which are related to the Keplerian elements through the inverse of Eqs.(\ref{eq: delaunay}). The equations of motion are then: 
\begin{eqnarray}\label{eq: eqmodelaunay}
    &~&\dot{\ell}=~~{\partial\mathcal{H}_{SM}\over\partial L},~~~~~~~~~~
    \dot{g}=~~{\partial\mathcal{H}_{SM}\over\partial G},~~
    \dot{h}=~~{\partial\mathcal{H}_{SM}\over\partial H},~~\nonumber\\
    &~&\dot{L}=-{\partial\mathcal{H}_{SM}\over\partial \ell}=0,~~
    \dot{G}=-{\partial\mathcal{H}_{SM}\over\partial g},~~
    \dot{H}=-{\partial\mathcal{H}_{SM}\over\partial h}~~.
\end{eqnarray}     
For numerical orbits, the above equations are integrated using the integrator proposed in \cite{10.1093/mnras/stab1032}, yielding the evolution of all six canonical variables, then transformed in the values of the six osculating Kaplerian elements. We call the orbits obtained in this way the `SELENA-Model' (SM) orbits.  

An important remark in what follows stems from the fact that, in the canonical formalism, the position-momenta variables correspond to orbital elements in a \textit{rest frame} whose axes instantaneously coincide with the orientation of the PALRF axes at time $t$. This implies that even for a pure Keplerian ellipse, i.e., had we set all perturbations $V_L, V_E, V_S$ equal to zero in Eq.(\ref{eq: hamsm}), the centrifugal term $-\omega_z H$ has the effect that the line of nodes, which is fixed in the rest frame, would rotate clockwise at a rate $\dot{h}=\partial\mathcal{H_{SM}}/\partial H = -\omega_z$. In reality, $\dot{h}$ is slightly different from minus the angular frequency of the Moon's revolution ($\simeq \omega_z$) because of the secular effects on the orbit caused by all three perturbations $V_L, V_E, V_S$. On the other hand, $\dot{g}$ is not influenced by considering the motion in a rotating frame, since the angle $g=\omega$ is always relative to the position of the line of nodes. This means that $\dot{g}=0$ for a fixed Keplerian ellipse, implying that, when all the perturbations are taken into account, the correct secular frequencies to compare in the PALRF frame are $\dot{g}$ and $\dot{h}+\omega_z$. The fact that the obliquity of the Moon is small, together with the presence of strong forces due to the lunar potential's zonal multipole harmonics, leads to a radically different structure of the web of secular resonances for the Moon's satellites compared to the case of Earth satellites, as discussed in detail in the next section.

\subsection{Numerical lifetime maps}
\label{ssec: numlife}
\begin{figure}
    \centering
    \includegraphics[width = \textwidth]{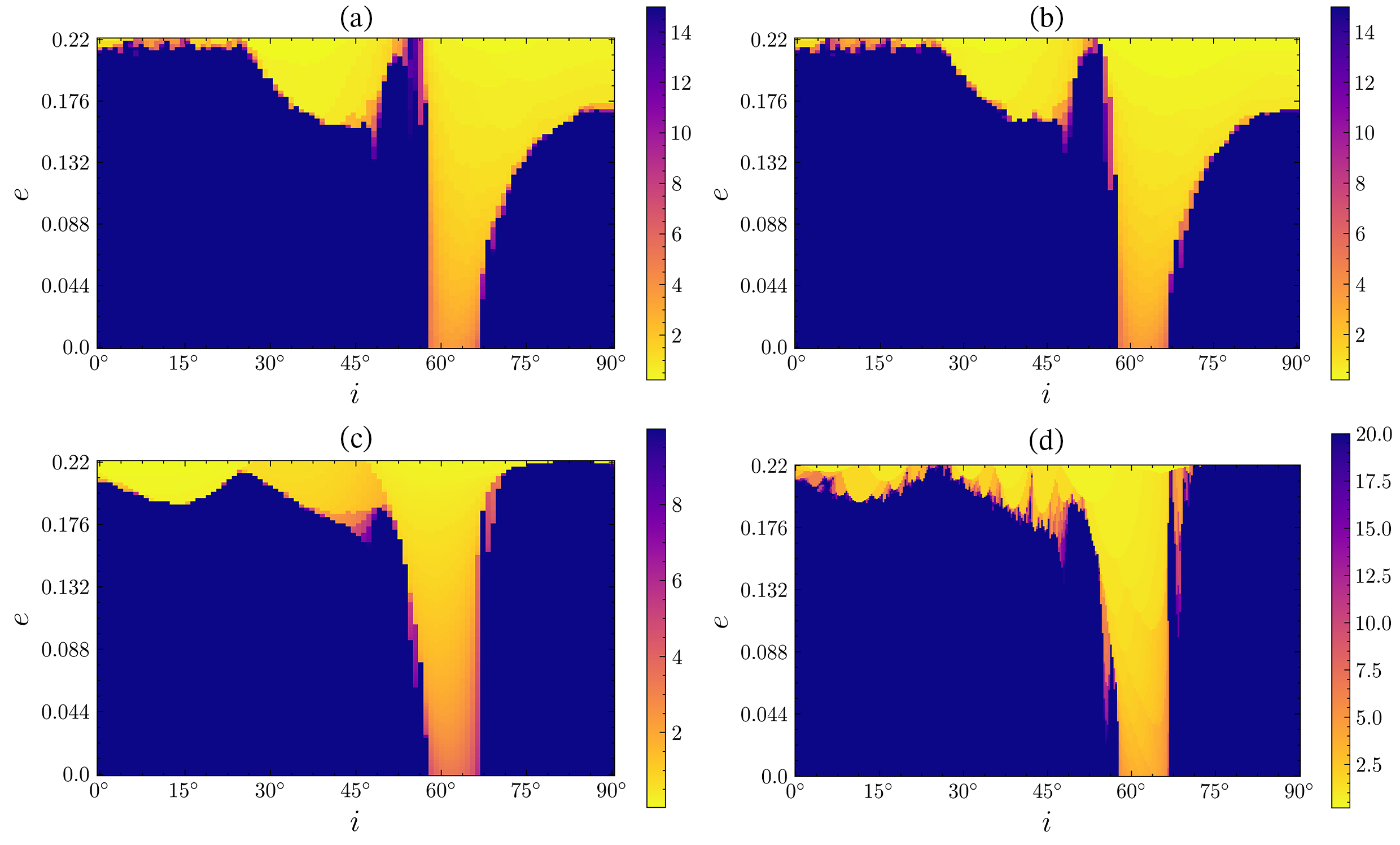}
    \caption{Four different realizations of the lifetime map for orbits with $a=r_L+500 km$. The cartography parameters are: 
    (top left) $n_g=9$, $N_g=100$, $\Omega_0=0^\circ$, $\omega_0=90^\circ$, $T_{run}=15$~yr. 
    (top right) $n_g=9$, $N_g=100$, $\Omega_0=20^\circ$, $\omega_0=70^\circ$, $T_{run}=15$~yr. 
    (bottom left) $n_g=10$, $N_g=100$, $\Omega_0=0^\circ$, $\omega_0=0^\circ$, $T_{run}=10$~yr. 
    (bottom right) $n_g=9$, $N_g=300$, $\Omega_0=0^\circ$, $\omega_0=90^\circ$, $T_{run}=20$~yr. }
    \label{fig:500kmlifetime}
\end{figure}
\begin{figure}
    \centering
    \includegraphics[width = \textwidth]{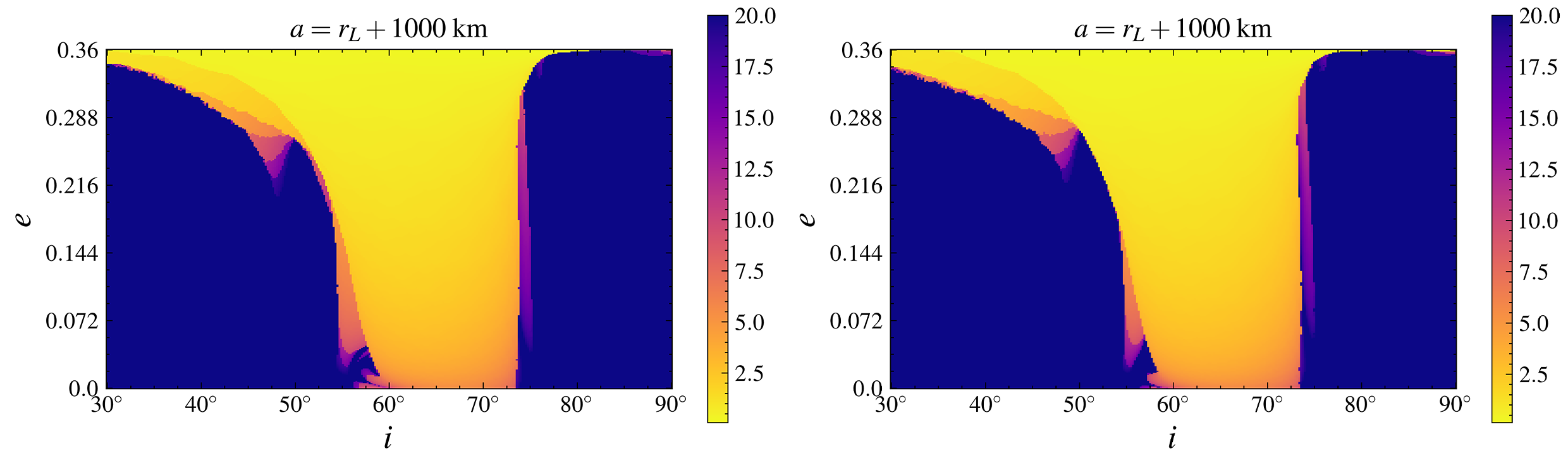}
    \caption{Lifetime cartography maps with $a=r_L+1000 km$ with $N_g=300$, $\Omega_0=0^\circ$, $\omega_0=0^\circ$, $T_{run}=20$~yr and $n_g=7$ (left),  $n_g=10$ (right). }
    \label{fig:1000kmlifetime}
\end{figure}
\begin{figure}
    \centering
    \includegraphics[width = \textwidth]{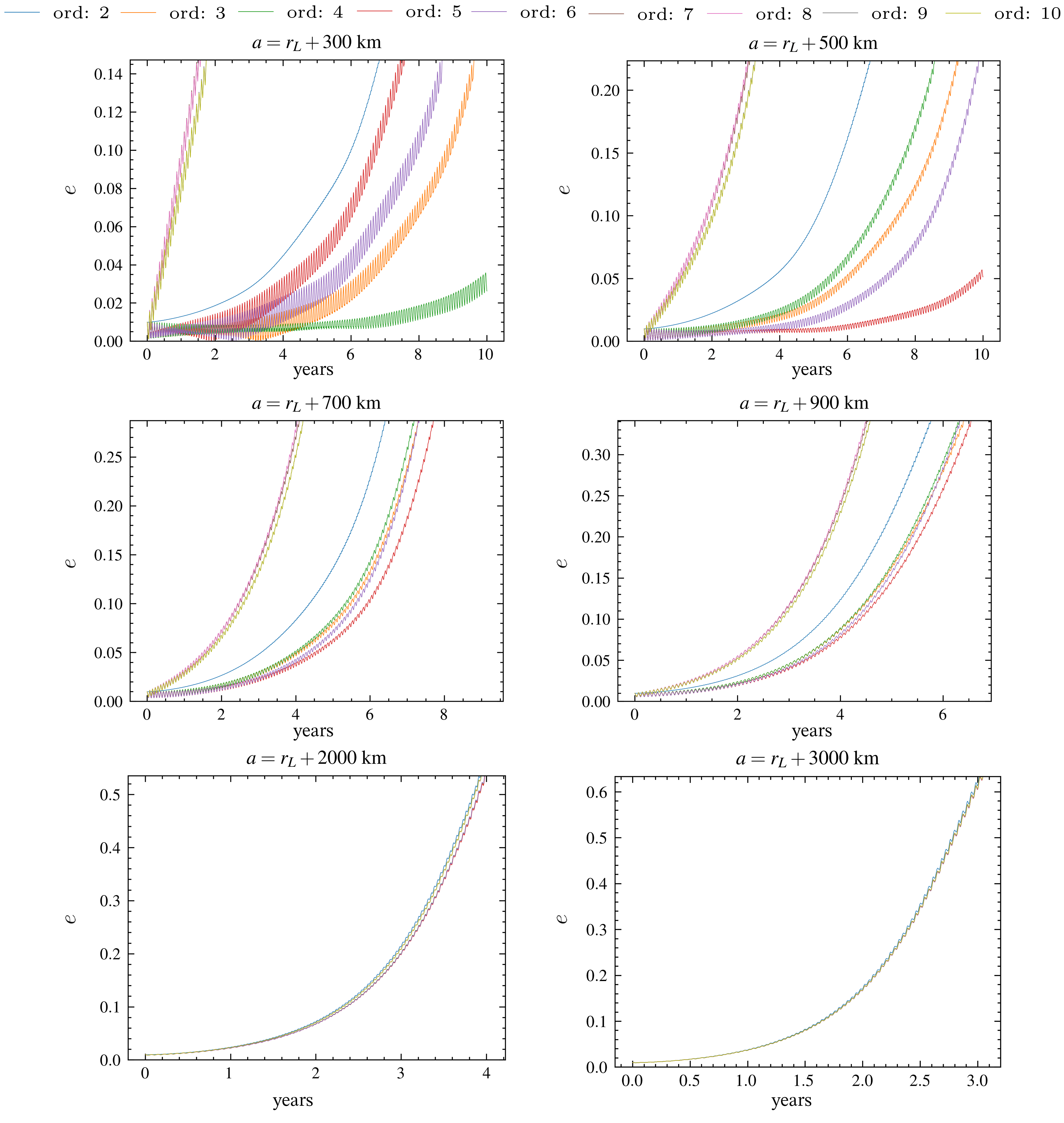}
    \caption{Each panel shows nine different realizations of the integration of one orbit with the lunar potential truncated at different maximum degrees varying from $n_g=2$ to $n_g=10$ (the resulting orbits are shown with different colors). In all six panels we have the same initial conditions for the five Keplerian elements $e_0=0, i_0=63.5^\circ, \Omega_0 = \omega_0 = m_0 = 0$, and semi-major axes corresponding to the altitudes of $300$ km, $500$ km, $700$ km, $900$ km, $2000$ km and $3000$ km. In all computed trajectories the perturbations due to the Earth and the Sun are included.
    }
    \label{fig:single_traj_comparison}
\end{figure}

\begin{figure}
    \centering
    \includegraphics[width = \textwidth]{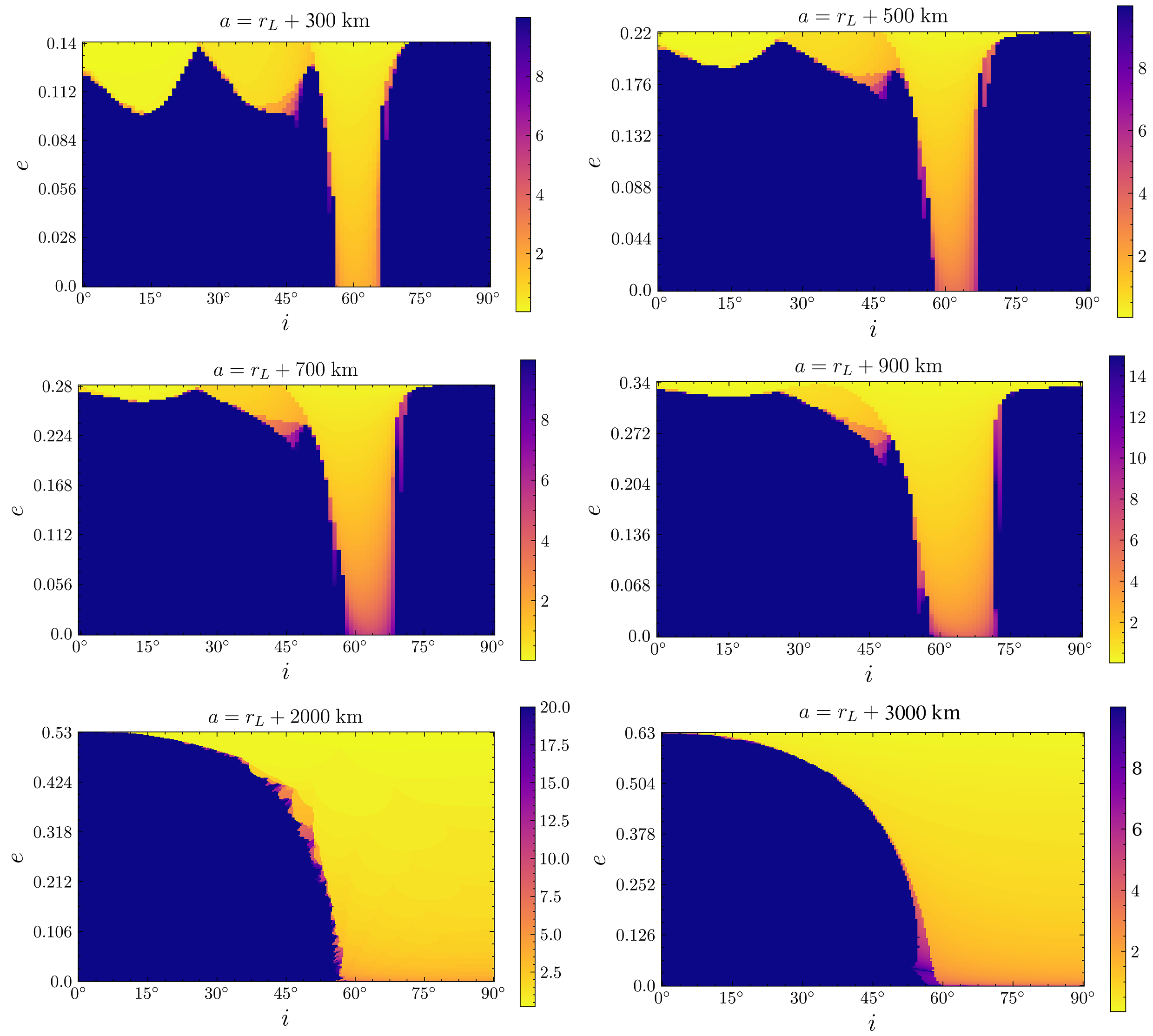}
    \caption{Lifetime maps with $n_g=10$, $N_g=100$, $\omega_0=\Omega_0=0$, $T_{run}$ as indicated in the top of the color bar in each panel, and various values of the altitude, as indicated in the legend above each panel. }
    \label{fig:300to3000kmlifetime}
\end{figure}
We define a satellite's \textit{orbital lifetime} as the time up to which a satellite, subject to secular variations of its orbital eccentricity, remains with a perigee satisfying the condition $a(1-e)>r_L$. Lifetime cartography at various `altitudes' (fixed values of the semi-major axis $a>r_L$) is obtained as follows: we consider a $N_g\times N_g$ rectangular grid of initial conditions for prograde orbits in the square $0\leq i(0)< 90^\circ$, $0\leq e(0)<e_{re}(a)$ where the re-entry eccentricity  is: 
\begin{equation}\label{eq: ere}
    e_{re} = 1 - r_L/a~~.
\end{equation}
We consider a fixed value of the angles $\ell(0)=M_0$, $g(0)=\omega_0$, $h(0)=\Omega_0$, same for all the points $i(0),e(0)$ in the grid, and run the numerical orbits as specified in the previous subsection up to a maximum time $T_{run}$. The orbital lifetime of a satellite is defined as the time $T_{re}$ of first occurence of its eccentricity becoming equal to $e_{re}$. If no re-entry takes place up to the end of the integration we set $T_{re}=T_{run}$. We then show in color map the function $T_{re}(i(0),e(0))$ as computed at all the points of the grid. In the numerical integration we test how the lifetime map is altered by including the lunar potential harmonics starting from degree $n=2$ and up to a maximum degree $n_g$, for various choices of $n_g$. 

Figure \ref{fig:500kmlifetime} shows lifetime maps computed for $a=r_L+500~$km. 
The four panels represent four different choices of the cartography parameters $(N_g,n_g,M_0,\omega_0,\Omega_0)$ as indicated in the caption. We notice that the structure of the lifetime maps varies marginally with the choice of degree $n_g$ close to the limiting value $n_g=10$, while no substantial detail is added to the map also when passing from a resolution $N_g=100$ to $N_g=300$. The only noticeable difference refers to orbits near the upper limit in both the initial inclination and the initial eccentricity (top right corner in the lifetime maps). We find that several orbits in this domain are protected from re-entry when $\omega_0$ is close to zero, while they evolve towards re-entry if $\omega_0$ is substantially larger than zero. This tendency will be explained in section 3 below.  

As the altitude is increased, in accordance with Figure \ref{fig:moon_force} we find that the lifetime map stabilizes after the inclusion of harmonics of lower maximum degree $n_g$. This is exemplified in Figure \ref{fig:1000kmlifetime}, where we see that truncating the force model at $n_g=7$ or $n_g=9$ yields practically equivalent lifetime maps. 
This is confirmed also by Figure \ref{fig:single_traj_comparison}, where a single orbit, with the same initial condition, is propagated using nine different  truncations of the lunar potential, i.e., at $n_g = 2, \dots, 10$. We see that the maximum degree $n_g$ after which the integration stabilizes to practically the same orbit decreases as the altitude increases, so that $n_g= 10$ is required at the altitude $a-r_L=300~$km, while $n_g=2$ is sufficient at $a-r_L=3000~$km. 

Following these numerical tests, Figure \ref{fig:300to3000kmlifetime} focuses on the main information obtained through the above-realized numerical lifetime cartography. The figure shows the lifetime maps obtained with the cartography parameters as indicated in the caption, for six different altitudes, namely $a=r_L+300$, $500$, $700$, $900$, $2000$, and $3000$~km. At lower altitudes ($a-r_L<1000$~km) we observe some sub-structure in the lifetime maps appearing in the form of some local minima of the border curve $e(i)$ separating initial conditions leading to re-entry from those which do not. In the next section, we will argue that such minima are connected to some low-order secular resonances besides the `$2g$' resonance (see next section). 
However, their presence in the lifetime maps is no longer distinguished at altitudes $a-r_L>1000$~km. Instead, from those altitudes on, we observe the presence of a nearly connected domain of initial conditions leading to re-entry, which has a nearly vertical right border and a nearly smooth concave left border. At the base of each map (for $e(0)=0$) both borders terminate at two limiting values of the inclination $(i_{min},i_{max})$. The left limit $i_{min}$ shows little variation with the altitude, around values $55^\circ<i_{min}<60^\circ$, while the right limit $i_{max}$ increases with the altitude, reaching $i_{max}=90^\circ$ at about the altitude $a-r_L=1300$~km. 

\section{Analytic Model}
\label{sec: analytic}
In the present section, we demonstrate that most features observed in the numerical lifetime maps, as exposed in the previous section, can be reproduced qualitatively, and to a large extent also quantitatively, through a simple model of the most important secular resonance of the problem, namely the $2g$ resonance. We first give some general account of the structure of secular resonances for lunar satellites, and then proceed to the interpretation of the lifetime maps through the phase portraits of the $2g$ resonance. 

\subsection{The web of secular resonances}
\label{ssec: webres}
The secular model $\mathcal{H}_{SM}$ involves five different frequencies whose commensurabilities potentially lead to resonance effects: $\dot{g}$ (precession of the satellite's perilune), $\dot{h}+\omega_z$ (precession of the satellite's line of nodes in a fixed frame coinciding with the PALRF frame at $t=0$)  $\nu_L,\dot{g}_L$ (precessions of the Moon 's line of nodes in the ecliptic plane, and the Moon's argument of the perigee). These frequencies enter into the Earth's radius vector $\vec{r}_E(t)$, and $\dot{\lambda_S}$ (the Sun's orbital frequency, which enters in the Sun's radius vector $\vec{r}_S(t)$). Up to five significant figures we have that $\omega_z(t)=const=2.64\times 10^{-6} ~rad/sec$ (corresponding to the Moon’s siderial lunar day equal to $27.55$ days), $\nu_L=1.07\times 10^{-8} ~rad/sec$ (corresponding to a period of $18.6$~yr), $\dot{g}_L=2.25\times 10^{-8} ~rad/sec$ (corresponding to a period of $8.85$~yr), $\dot{\lambda}_S=1.99\times 10^{-7}~rad/sec$. However, from Fig.\ref{fig:moon_force} we deduce that the solar force is very small at all altitudes here considered, while the frequency 
$\dot{g}_L$ only involves terms of small size depending on the eccentricity of the Moon's geocentric orbit ($e_L=0.054$) and can also be ignored. On the other hand, in secular theory the frequencies $\dot{g}$ and $\dot{h}$ are functions of the elements $(a,e,i)$ and can be computed as follows: first, from the tables provided in \cite{efthymiopoulos2023selena}, we keep the most important terms in the trigonometric representation of the Earth and Sun PALRF vectors, assuming i) circular orbits for both bodies, and ii) a zero obliquity of the Moon with respect to the ecliptic (the true obliquity is equal to $1.5^\circ$). Finally, we set the Earth's position to depend only on one angle $\tau_E=\omega_z t$ with period the sidereal lunar day 
\begin{align} \label{eq: earthvec}
    x_E(t)[km] &= 382470 + 14800 (\cos\tau_E + \sin\tau_E), \nonumber\\
    y_E(t)[km] &= 29750 ( \cos\tau_E - \sin\tau_E), \\
    z_E(t)[km] &= -44650  \cos\tau_E, \nonumber
\end{align}
and analogously the Sun's position to depend only on one angle $\tau_S=(\omega_z-\dot{\lambda}_S) t$ with period the synodic lunar day (29.53 days)
\begin{align}\label{eq: sunvec}
    x_S(t)[km] &=  - 6.9917\times 10^7 \cos\tau_S - 1.322\times 10^8 \sin\tau_S, \nonumber\\
    y_S(t)[km] &= -1.322\times 10^8  \cos\tau_S + 6.9917\times 10^7 \sin\tau_S, \\
    z_S(t)[km] &= 0~. \nonumber
\end{align}
Substituting the above expressions in the Hamiltonian $\mathcal{H}_{SM}$, setting $x_E(t)=a_L+\Delta x_E(t)$, with $a_L=382470$~km, expanding up to first order in the three small quantities $\Delta x_E(t)$, $y_E(t),z_E(t)$, noticing that only the constant Sun's distance $r_S^3=(x_S(t)^2+y_S(t)^2)^{3/2}$ appears in the denominator of the $V_S$, and, finally, making all trigonometric reductions, we obtain an approximate model for the Hamiltonian which takes the form:
\begin{equation}\label{hsminteplus}
\mathcal{H}_{SM}=
\mathcal{H}_{SM,0}(a,e,i)+\mathcal{H}_{SM,1}(a,e,i,g,h,\tau_E,\tau_S)~~.
\end{equation}
Then, we compute:
\begin{align}\label{ghdot}
    \dot{g}(a,e,i) &= {\partial\mathcal{H}_{SM,0}\over\partial G} 
    = 
    {1\over\sqrt{\mu_La}}\left({\eta\over e}{\partial\mathcal{H}_{SM,0}\over\partial e}
    -
    {\cos i\over\eta\sin i}{\partial\mathcal{H}_{SM,0}\over\partial i} \right)  \\
    \dot{h}(a,e,i) &= {\partial\mathcal{H}_{SM,0}\over\partial H} 
    = 
    {1\over\sqrt{\mu_La}}\left({1\over\eta\sin i}{\partial\mathcal{H}_{SM,0}\over\partial i} \right)~.  \nonumber
\end{align}
Note that, besides $(g,h,\tau_E,\tau_S)$, in the complete SM the Hamiltonian $\textit{H}_{SM,1}$ depends also on the angles $\nu_L t$ and $\dot{g}_L t$, the latter dependence being, however, through terms of negligible size. 

We then define the `$k_1 g +k_2(h+\lambda_L)+k_3\Omega_L$ resonance', with $\vec{k} = \left(k_{1}, k_{2}, k_{3}\right)$  $\in \mathbb{Z}^{3}$ as the two dimensional surface $\mathcal{R}_{\vec{k}}$ in the 3D space of elements $(a,e,i)$ where the commensurability 
\begin{equation}
    \label{eq: resonance}
    \mathcal{S}_{\vec{k}}:=\left\{(a,e,i):~k_{1}\dot{g}(a,e,i) +k_{2} (\dot{h}(a,e,i)+\omega_z)+k_{3}\nu_L=0\right\}~
\end{equation}
holds. Here, the angle $\lambda_L$ is defined as $\lambda_L=\lambda_0+\tau_E$, and it indicates the angle of the PALRF $x-$axis with respect to a chosen fixed frame. The value of $\lambda_0$ depends on the choice of fixed frame and on the initial epoch.

\begin{figure}
    \centering
    \includegraphics[width = \textwidth]{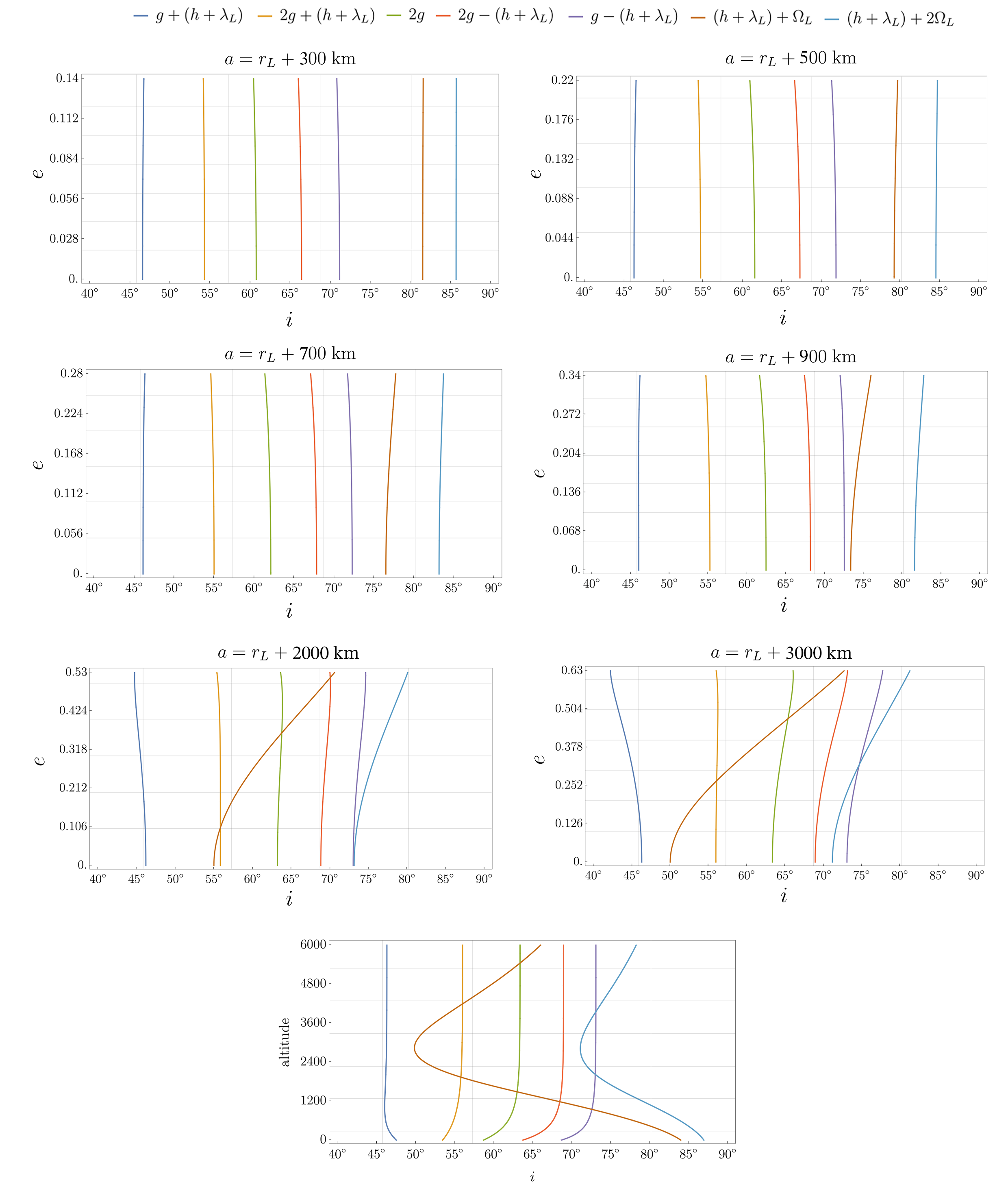}
    \caption{The curves corresponding to the intersection of the resonant surfaces $\mathcal{R}_{\vec{k}}$ with the plane $a=const$, with the constant value of the semi-major axis as indicated in each panel, for the most important secular resonances of the problem. The altitudes are the same as in the corresponding lifetime maps of Fig.\ref{fig:300to3000kmlifetime}. }
    \label{fig:300to3000kmresonances}
\end{figure}
Figure \ref{fig:300to3000kmresonances} shows the curves corresponding to the most important (low-order) resonant surfaces $\mathcal{R}_{\vec{k}}$ in the $(i,e)$ plane for fixed $a$ and at the bottom in the $(i,a)$ one for fixed $e=0$. Opposing Fig.\ref{fig:300to3000kmresonances} to Fig.\ref{fig:300to3000kmlifetime}, we can observe that, at all altitudes, the domains where re-entry takes place are crossed by several of the resonances shown in Fig.\ref{fig:300to3000kmresonances}. 
However, a quick analysis can show that, at both low and high altitudes, not all of these resonances are important for eccentricity growth. Deferring a detailed analysis to a separate paper, we may summarize the way resonances act in this problem as follows. Recalling basic symmetries (the so-called `D'Alembert' rules) of the Hamiltonian, it follows that:
\begin{enumerate}
\item The Hamiltonian $\mathcal{H}_{SM,0}$ contains contributions only from the even degree zonal harmonics ($J_2, J_4,\ldots$) of the lunar multipole potential expansion $V_L$, as well as a negligible contribution from the Earth's potential term $V_E$ (of relative importance $\mathcal{O}((\mu_E/\mu_L)(a/R_L)^2$ $(a/a_L)^3)$ with respect to the $J_2$ term, see Fig.\ref{fig:moon_force}). Thus, the locations of all resonances are determined, practically, by the even zonal harmonics $J_{2j}$, $j=1,2,\ldots$. We refer to this property in short as that \textit{`the even zonal harmonics define the centers of the resonances'}.
\item The $J_2$ term contributes no trigonometric term of first order in $J_2$ to the Hamiltonian term $\mathcal{H}_{SM,1}$, while it contributes a (here neglected) term $\mathcal{O}(J_2^2)\cos(2g)$ after second order normalization in closed form. The remaining even zonal harmonics $2j>2$ all contribute a first order $\mathcal{O}(J_{2j}R_L^{2j}e^2/a^{2j+1})\cos(2g)$ term to $\mathcal{H}_{SM,1}$.  On the other hand, the odd zonal harmonics $J_{2j+1}$, $j=1,2,\ldots$ contribute to odd trigonometric terms $e\sin(g), e^3\sin(3g)$, etc, with amplitudes following a similar scaling as for the odd zonal terms. The Earth's potential contributes a $\mathcal{O}(e^2\mu_E (a/a_L)^3)\cos(2g)$ to $\mathcal{H}_{SM,1}$. We shortly refer to the above set of properties as that \textit{`the exact position of the fixed points, and the form of the separatrices of the $2g$-resonance, are determined by both the zonal harmonics $J_n$, $n=2,3,\ldots$ and the Earth's tidal term $V_E$'}. As seen in Fig.\ref{fig:moon_force}, owing to their large value (due to the lunar mascons), the zonal harmonics are the main terms to determine the position of the fixed points and form of the separatrices of the $2g-$resonance, at least up to the end of the middle-altitude zone.
\item No zonal harmonic contributes to any other secular resonance of the ensemble (\ref{eq: resonance}). Tesseral harmonics $C_{n,m}$, $S_{n,m}$, with $n\geq 2$, $1\leq m\leq n$, do not contribute any term to $\mathcal{H}_{SM,0}$, while they contribute cosine and sine trigonometric terms to $\mathcal{H}_{SM,1}$. The latter terms necessarily contain the satellite's longitude of the nodes $h$, but not $h+\lambda_L$, which is the argument of interest for secular resonances. On the other hand, trigonometric terms involving the argument $h+\lambda_L$ are due to the Earth's potential $V_E$, of relative importance proportional to the sine of the Moon's obliquity $\varepsilon_L\simeq 0.12~rad$. We shortly refer to this property as that \textit{`the form of the separatrices of all secular resonances of the form $k_1g+k_2(h+\lambda_L)$ with $k_2\neq 0$ is determined only by terms in $V_E$ proportional to $\sin(\varepsilon_L)$'}. Since such terms are one order of magnitude smaller than the respective Earth's contribution to zonal harmonics, their impact on the structure of the resonance web is negligible even for high-altitude orbits. 
\item Trigonometric terms in $\mathcal{H}_{SM,1}$ depending on the angle $\Omega_L$ are produced only by the Earth potential $V_E$, and they are of order $\mathcal{O}(e^2\mu_E (a/a_L)^3)$ $\sin(\varepsilon_L)\sin(i_p)$, where $i_L$ is the inclination of the Moon's orbit with respect to the ecliptic, $i_L\simeq 0.09~rad$. 
\end{enumerate}    

\subsection{Simplified Hamiltonian model}
\label{ssec: hamssm}
\begin{figure}
    \centering
    \includegraphics[width = \textwidth]{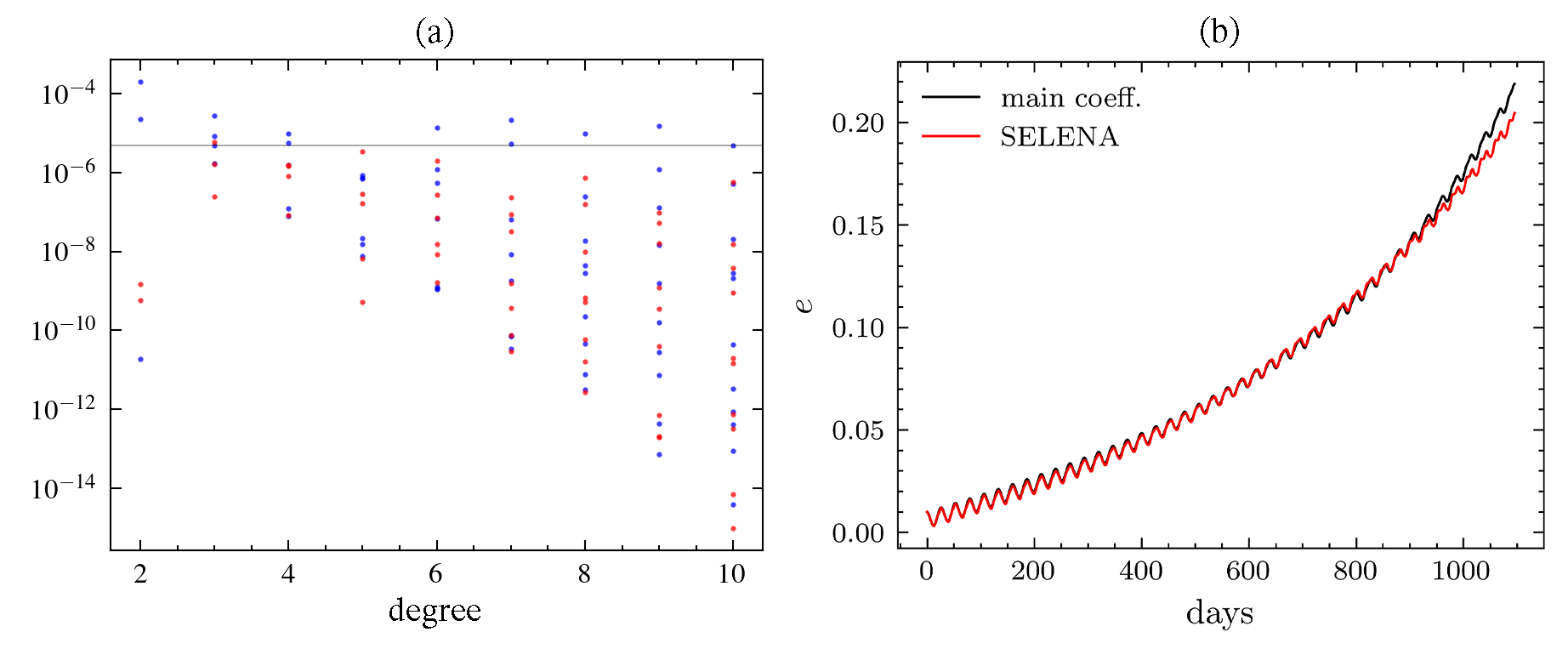}
    \caption{
        Left: The values of the normalized coefficients $\bar{C}_{nm}$ (blue) and $\bar{S}_{nm}$ (red) as a function of the order $n$. The horizontal line corresponds to the threshold value $5\times 10^{-6}$. Numerical values are reported in \ref{app: GRAIL}. Right: Comparison between the evolution of the eccentricity of an orbit with initial conditions $a=r_L + 500$ km, $e = 0.01$, $i = 60^\circ$, $\Omega = 0^\circ$, $\omega = 0^\circ$, $M = 0^\circ$ using the complete Hamiltonian $\mathcal{H}_{SM}$ (red), or the simplified model $\mathcal{H}_{SSM}$ (black).
        }
    \label{fig:simplemodel}
\end{figure}
Figure \ref{fig:simplemodel}, in conjunction with what was exposed in the previous subsection, allows now to construct a `simplified SELENA-mean' secular Hamiltonian model (hereafter $\mathcal{H}_{SSM}$), which contains only a small subset of terms of $\mathcal{H}_{SM}$, while, yielding essentially the same long-term behavior of the orbits. The left panel of Fig.\ref{fig:simplemodel} shows the values of all the normalized GRAIL coefficients $\bar{C}_{nm}$ (blue) and $\bar{S}_{nm}$ (red) as a function of the degree $n$, with $2\leq n\leq 10$. These values are reported in Appendix I. The horizontal line corresponds to an arbitrary threshold value $5\times 10^{-6}$. We find twelve harmonics whose normalized value is above such threshold, namely the set 
\begin{equation}\label{eq:ssmset}
    \mathcal{CS}_{SSM}=\left\{C_{20}, C_{22}, C_{30}, C_{31}, S_{31}, 
    C_{40}, C_{41}, C_{60},C_{70}, C_{71}, C_{80}, C_{90}\right\}~.
\end{equation}
Using i) only the harmonics of Eq.(\ref{eq:ssmset}), ii) the simplified form of the PALRF Earth vector, given in Eq.(\ref{eq: earthvec}) (expanding again up to first order with respect to the small quantities $\Delta x_E(t)$, $y_E(t),z_E(t)$, iii) ignoring the negligible contribution of the Sun  and iv) setting $\omega_z=const=\omega_L=2.64\times 10^{-6} rad/sec$, as well as $\omega_x=\omega_y=0$, leads to the \textit{simplified model} $\mathcal{H}_{SSM}$. The right panel of Fig.\ref{fig:simplemodel} shows an orbit leading to re-entry, as integrated with the full Hamiltonian $\mathcal{H}_{SM}$ of the SELENA-mean model, or with the simplified Hamiltonian $\mathcal{H}_{SSM}$, showing that the orbital evolution is nearly identical with the two models all the way up to the re-entry. 

\begin{figure}
    \centering
    \includegraphics[width = \textwidth]{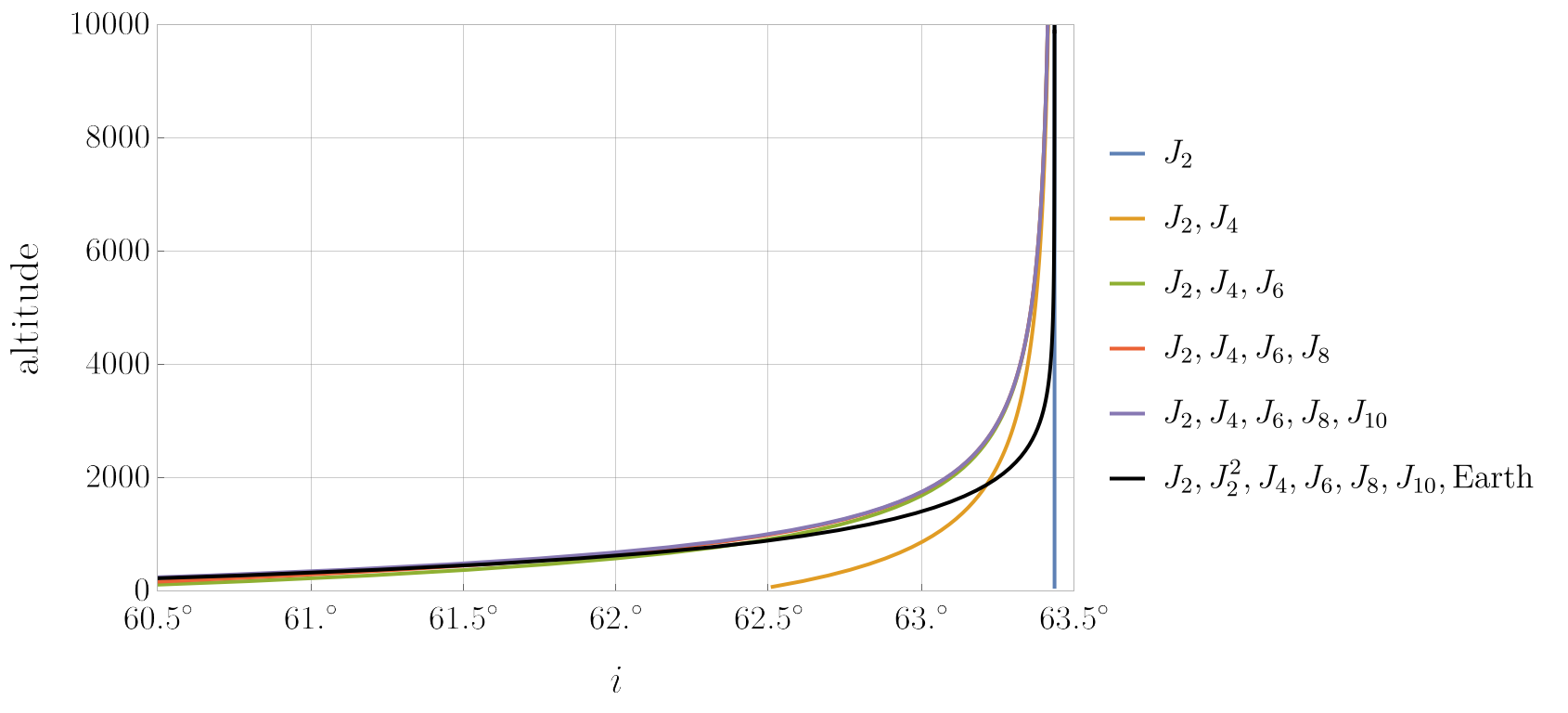}
    \caption{The inclination at which $\dot{\omega} = 0$ for circular orbits as a function of the altitude.}
    \label{fig:location_i2g}
\end{figure}
As an additional test, Fig.\ref{fig:location_i2g} shows the value of the inclination $i(a)$ where the $2g-$resonance (plane $\mathcal{S}_{2,0,0}$) intersects the plane $e=0$, as a function of the altitude $a-r_L$. The black curve shows the computation with the integrable part of the complete model $\mathcal{H}_{SM,0}$ plus the $J_2^2$ terms of the SELENA model. The remaining curves show the same curve as obtained under various truncations in the Moon's zonal harmonics. Note that the $J_2$ term alone yields $i_{critical}=63.4^\circ$, which holds as well for any value of $e\neq 0$. Since the $J_2$ term is dominant over all other lunar harmonics at high altitude, the location of the resonance converges towards the value $i_{crit}$ as $a-r_L$ increases, and the critical value is essentially reached at $a-r_L\approx 5000$~km. As concluded from the convergence of the curves, the correct dependence of the location of the $2g-$resonance on the altitude is essentially recovered for all altitudes $a-r_L>100$~km, by the addition of the first three important zonal harmonics, i.e., adding $J_2$, $J_4$ and $J_6$.

\begin{figure}
    \centering
    \includegraphics[width = \textwidth]{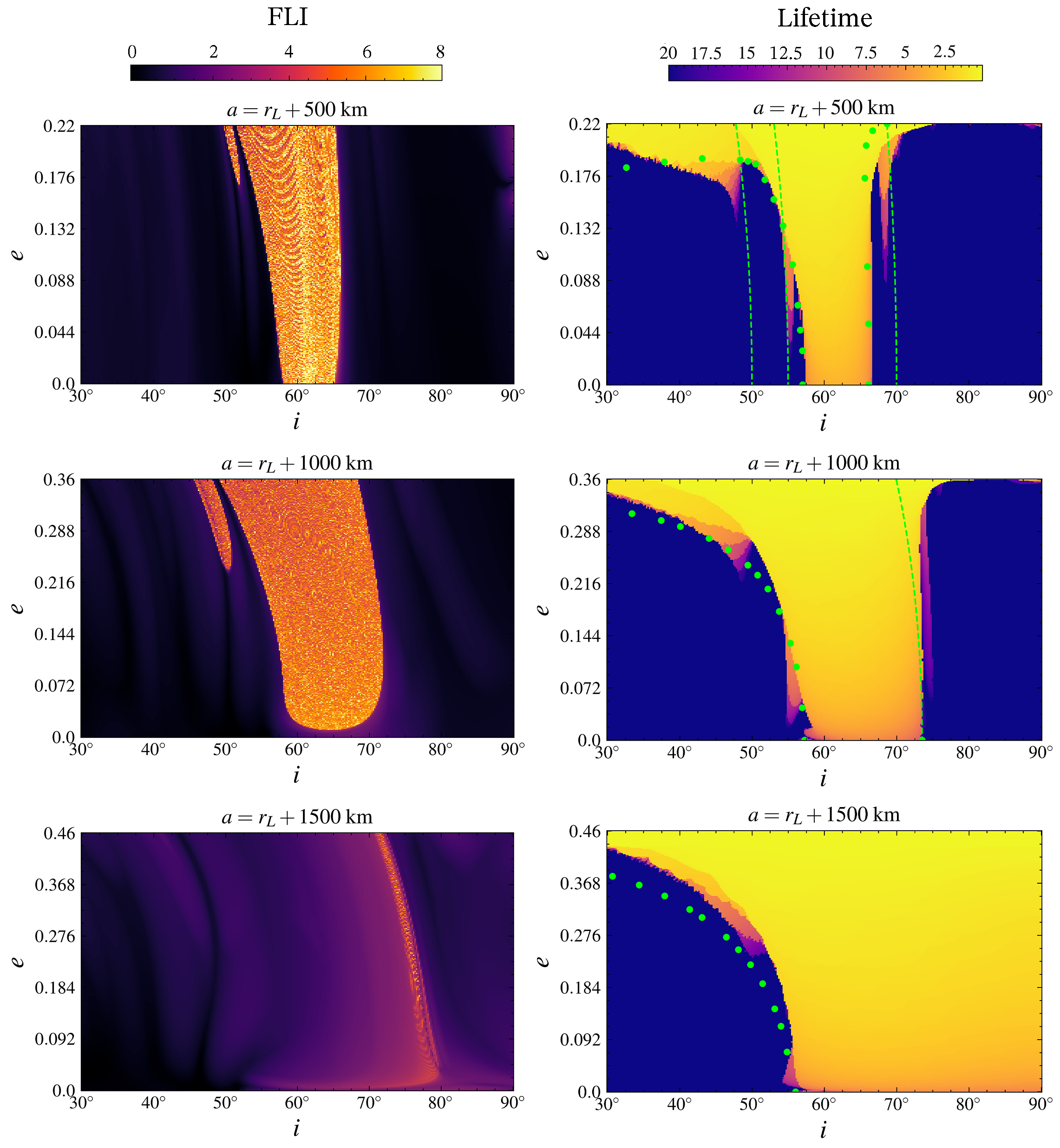}
    \caption{(left) FLI maps, and (right) lifetime maps in a grid of initial conditions in the $(i,e)$ plane for orbits with $\Omega = \omega=0^\circ$ at the altitudes $a-r_L=500 km$ (top row), $a-r_L=1000 km$ (middle row), $a-r_L=1500 km$ (bottom row). The numerical trajectories are computed in the complete Hamiltonian $\mathcal{H}_{SM}$, while the FLI is computed by calculating the associated variational equations of motion. The thick points in the lifetime maps correspond to theoretical predictions of the borders of the domains of re-entry obtained through the model of the $2g-$resonance discussed in subsection \ref{ssec:2gres}. The dashed curves indicate the `planes of fast drift' defined in the same subsection. }
    \label{fig:LT_maps}
\end{figure}

Having checked that the terms of the ensemble $\mathcal{S}_{SSM}$ are the only important ones as regards the behavior of secular resonances, we proceed to check which of the resonances have well developed separatrices as a function of the altitude $a-r_L$. By property 2 above, all the zonal terms within the set $\mathcal{S}_{SSM}$ (even or odd) contribute only to the separatrices of the $2g$ resonances. On the other hand, by D'Alembert rules the tesseral harmonics $C_{22}$, $C_{31}$, $S_{31}$ $C_{41}$ and $C_{71}$ generate trigonometric terms according to the rule that the $C_{nm}$ or $S_{nm}$ harmonic generates terms with arguments $m_1g\pm m h$, where $0\leq m_1\leq n-2$. None of these terms involve the angle $h+\lambda_L$, thus the tesseral terms do not create separatrices for any of the secular resonances of the problem. Finally, the Earth's term contributes to the separatrices of all the secular resonances involving the angle $m(h+\lambda_L)$ through terms $\mathcal{O}(\mu_E a^2/a_L^3 \sin(m\varepsilon_L)$, where $\varepsilon_L \simeq 0.1$ is the obliquity of the Moon. We have already seen that the coefficient $\mu_E a^2/a_L^3$ yields a negligible contribution to the $2g$ resonance compared with the zonal harmonics of the lunar potential. Since the tesseral harmonics do not contribute to other resonances, and the Earth yields a negligible contribution, we conclude that \textit{no resonance other than $2g$ has developed separatrices in all three zones of altitude considered in the present study, i.e. up to $a-r_L=10000$~km.}

The above conclusion can be substantiated with numerical experiments in the complete model $\mathcal{H}_{SM}$, obtaining an indirect indication of the separatrix width of each of the resonances of Fig.\ref{fig:300to3000kmresonances} by computing Fast Lyapunov Indicator (FLI) maps (\cite{froeschle1997fast}), as in the left column of Fig.\ref{fig:LT_maps} for three different altitudes. The right column in the same figure shows the lifetime maps for the same grids of initial conditions, as indicated in the caption. The FLI maps reveal the existence of separatrix structures associated with the $2g$ resonance, but give no hint of important secular resonances other than $2g$. Comparing with the lifetime maps, however, reveals that the domains of initial conditions leading to eccentricity growth and re-entry are way more extended than the chaotic separatrix layers of the $2g$ resonance indicated through the FLI maps. 
We now theoretically interpret this effect, following an analysis of the structure of the separatrices of the $2g$ resonance in the simplified model $\mathcal{H}_{SSM}$.

\subsection{Integrable model and the separatrices of the  2g-resonance}
\label{ssec:2gres}
A resonant Hamiltonian model for the $2g$ resonance can be obtained starting from the simplified model $\mathcal{H}_{SSM}$, by following the same procedure as in \cite{Legnaro_Efthymiopoulos_2022} for the case of Earth satellites. We here summarize only the main steps leading to this model. 
First, we consider \emph{Modified Delaunay variables}
\begin{equation}
    \def\arraystretch{1.25}
    \centering
    \begin{tabular}{ l l }
    $\Lambda = L = \sqrt{\mu_L a} $,   &    $\lambda = l+g+h = M + \omega + \Omega$,  \\
    $P = L-G = L \left( 1-\sqrt{1-e^2}\right)$,   &    $ p = -g-h = -\omega - \Omega $,\\
    $Q = G - H = L\sqrt{1-e^2}(1-\cos i)$, $\quad$  &    $ q = -h = - \Omega. $\\
    \end{tabular}
    \label{eq: MOD_DEL_variables}
\end{equation}
We fix a value of the semimajor axis, which, in turn, gives the location $i_\star$ of the $2g$ resonance, according to the procedure discussed in the previous subsection. 
Next, we perform a series expansion in powers of the Hamiltonian $\mathcal{H}_{SSM}$ in powers of the small quantities $P$ (which is $\mathcal{O}(e^2)$) and $Q-Q_*$ (which is of the order $\mathcal{O}(i-i_*)$, with $Q_*=Q(e=0,i=i_*)$). This leads to a truncated polynomial representation of the Hamiltonian $\mathcal{H}_{SSM}$ in the variables $(P,\Delta Q=Q-Q_*)$. 
We then introduce \textit{resonant action-angle variables} 
\begin{equation}
    J_R = P, \quad J_F = P + (Q-Q_*), \quad u_R = p-q, u_F = q,
\end{equation}
and, finally, the \textit{Poincaré canonical variables}
\begin{equation}
    X = \sqrt{2 J_R} \sin u_R, \qquad Y = \sqrt{2 J_R} \cos u_R~~~.
\end{equation}
The Hamiltonian becomes now a function of the resonant variables 
\begin{equation}
    \mathcal{H}_{SSM}=\mathcal{H}_{SSM}(X,Y, J_F, u_F, \tau_E; a).
\end{equation}

To get an integrable model of the $2g$ resonance, we compute the average of $\mathcal{H}_{SSM}(X,Y, J_F, u_F, \tau_E; a)$ over the `fast' angles $(u_F,\tau_E)$ (faster than the resonant angle $g$). This is a polynomial in $X$ and $Y$, with $J_F$ and $a$ as parameters:
{\small \begin{align}
    \begin{split}
        &\mathcal{H}_{I}(X,Y;J_F,a) = \frac{1}{(2 \pi)^2} \int_{0}^{2 \pi} d u_F \int_{0}^{2 \pi}d \tau_E \; \; \mathcal{H}_{SSM}(X,Y, J_F, u_F, \tau_E; a) = \\
        &\omega_F J_F + \alpha^2 J_F^2 + \\
        & R_{10} X + M_{110} J_F X + M_{120} J_F^2 X + \\
        & R_{20} X^2 + M_{120} J_F X^2 + M_{220} J_F^2 X^2 + R_{02} Y^2 + M_{102} J_F Y^2 + M_{022} J_F^2 Y^2 + \\
        &  R_{30} X^3 + M_{130} J_F X^3 + M_{230} J_F^2 X^3 + R_{12} X Y^2  + M_{112} J_F X Y^2+ M_{212} J_F^2 X Y^2 + \\
        &R_{40} X^4 +R_{04} Y^4 + M_{140} J_F X^4 + M_{104} J_F Y^4 + R_{22} X^2 Y^2 + M_{122} J_F X^2 Y^2 + \\
        & R_{50} X^5 + R_{32} X^3 Y^2 + + M_{132} J_F X^3 Y^2 + R_{14} X Y^4 + M_{114} J_F X Y^4 + \\
        & R_{60} Y^6 + R_{42} X^4 Y^2 + R_{24} X^2 Y^4 + R_{52} X^5 Y^2 + \dots
    \end{split}
\end{align} }%
where $\omega_F=-\dot{h}(a,e=0,i=i_*)$. 

\begin{figure}
    \centering
    \includegraphics[width = 0.97\textwidth]{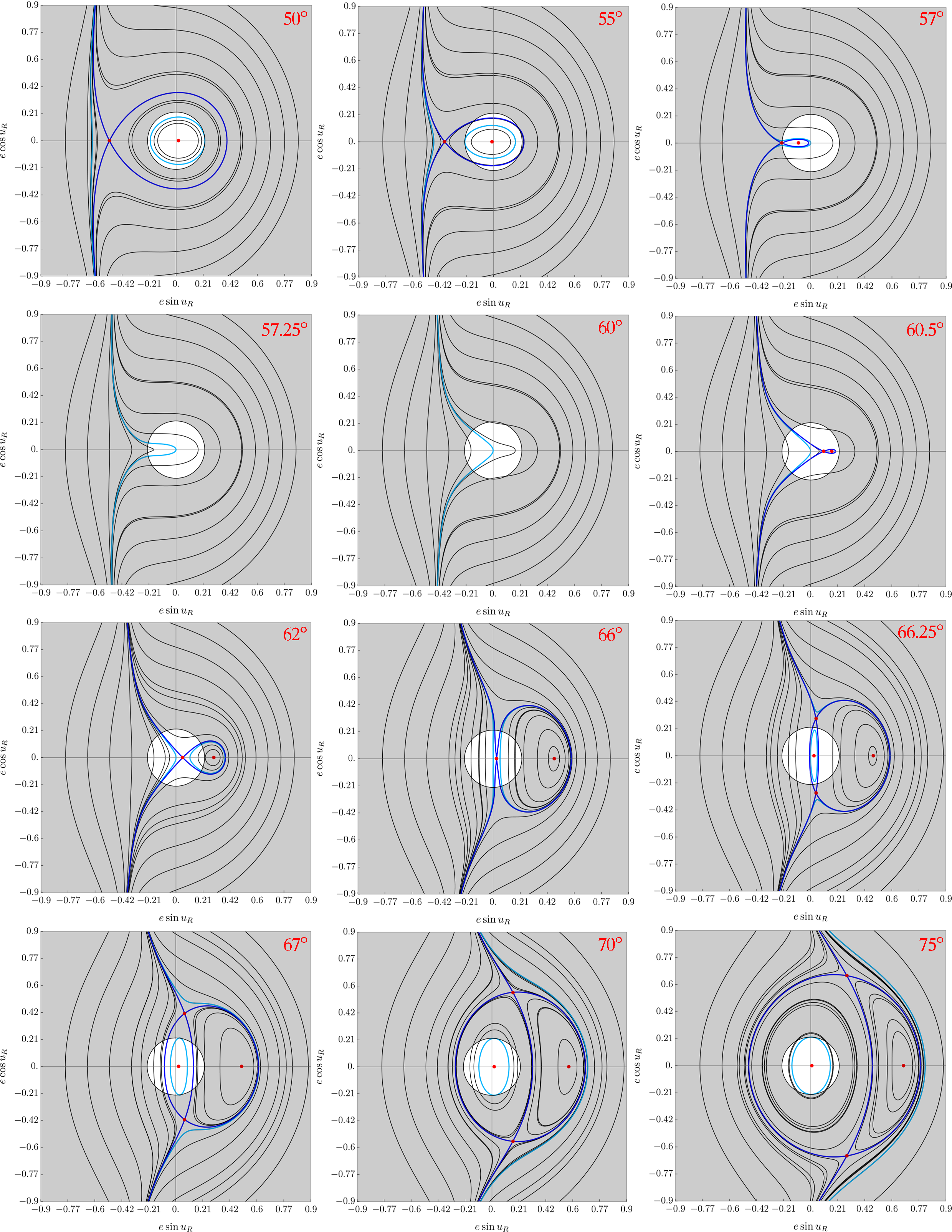}
    \caption{Phase portraits of the integrable Hamiltonian $\mathcal{H}_I$ at an altitude of $a-r_L=500$ km at the inclinations  $i_0=50 ^\circ$, $55 ^\circ$, $57 ^\circ$, $57.25 ^\circ$, $60 ^\circ$, $60.5 ^\circ$, $62 ^\circ$, $66 ^\circ$, $66.25 ^\circ$, $67 ^\circ$, $70 ^\circ$, $75 ^\circ$.
    The blue curve shows the separatrix passing through the central fixed point, whenever this point is unstable. Light-blue curves are tangent and entirely contained within the disc $e=e_{re}$ end define the borders of initial conditions of orbits protected from collision. Frozen orbits are plotted in red.}
    \label{fig:Phase_Portraits_500km}
\end{figure}
\begin{figure}
    \centering
    \includegraphics[width = 0.97\textwidth]{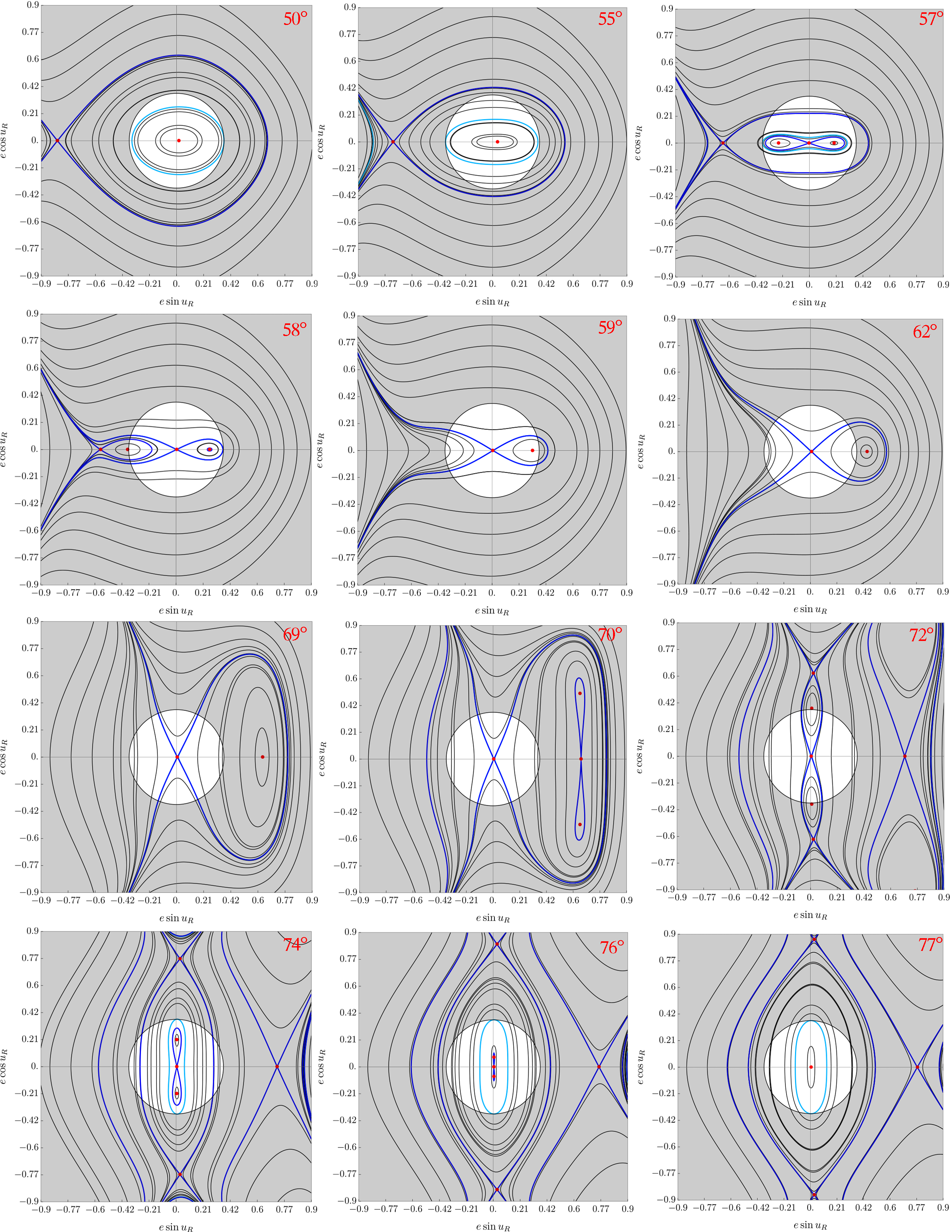}
    \caption{Same as in Fig.\ref{fig:Phase_Portraits_500km}, but for the altitude of $1000$ km at the inclinations  $i_0=50 ^\circ$, $55 ^\circ$, $57 ^\circ$, $58 ^\circ$, $59 ^\circ$, $62 ^\circ$, $69 ^\circ$, $70 ^\circ$, $72 ^\circ$, $74 ^\circ$, $76 ^\circ$, $77 ^\circ$.}
    \label{fig:Phase_Portraits_1000km}
\end{figure}
\begin{figure}
    \centering
    \includegraphics[width = 0.97\textwidth]{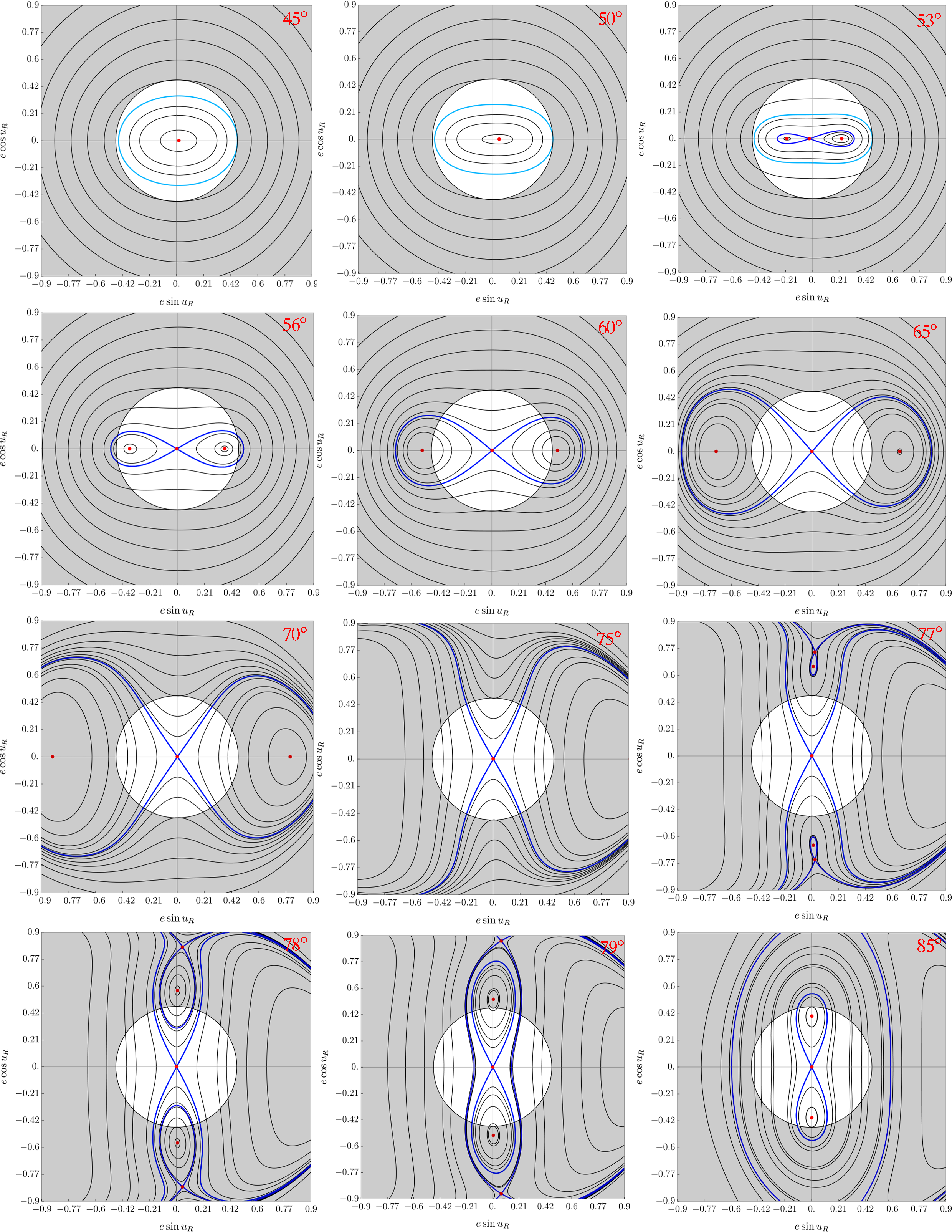}
    \caption{Same as in Fig.\ref{fig:Phase_Portraits_500km}, but for the altitude of $1500$ km at the inclinations  $i_0=45 ^\circ$, $50 ^\circ$, $53 ^\circ$, $56 ^\circ$, $60 ^\circ$, $65 ^\circ$, $70 ^\circ$, $75 ^\circ$, $77 ^\circ$, $78 ^\circ$, $79 ^\circ$, $85 ^\circ$.}
    \label{fig:Phase_Portraits_1500km}
\end{figure}

The above Hamiltonian is integrable, since $J_F$ is a second integral of motion in addition to the energy. From $\mathcal{H}_I$ it is possible to plot phase portraits in the $(X,Y)$ plane, which can then be transformed to plots in the plane $(e\sin(u_R),e\cos(u_R))$ through the relations $e=\sqrt{1-(1-(P/\sqrt{\mu_L a}))^2}$, with $P=(X^2+Y^2)^{1/2}$, and $u_R=-g=\arctan(X,Y)$. Figures \ref{fig:Phase_Portraits_500km}, \ref{fig:Phase_Portraits_1000km} and \ref{fig:Phase_Portraits_1500km} show a set of phase portraits obtained in the above way. Each portrait is parameterized by a constant value of $J_F$. The locus of points in the surface $(i,e)$ where 
\begin{equation}\label{eq:planefd}
\sqrt{\mu_L a}(1-\sqrt{1-e^2}\cos i)-Q_*=J_F
\end{equation}
is called \emph{fast drift plane}, and allows to map each point of the phase portrait to a point in the plane $(i,e)$, as explained in detail in \cite{daquin2021deep}. The inclinations shown as labels in each phase portrait in the above pictures, correpond to the inclination of a circular orbit with parameter equal to the label value $J_F$, $i_0=\cos^{-1}\left(1-((J_F+Q_*)/\sqrt{\mu_La})\right)$.

One important remark regarding these figures is that, while the shown portraits extend to eccentricities up to $0.9$, in reality the computation is valid only in the inner white disc domains whose border represents the collision condition $e=e_{re}(a)$. In fact, points exterior to the discs correspond to orbital radii smaller than $r_L$, hence, where the multipole expansion of the Moon's potential is no longer valid. However, we here show how these phase portraits formally look even ignoring the collision singularity of the multipole expansion, since this allows to visualize the domains where the separatrices formed by unstable fixed points \textit{within the discs} separate bounded from unbounded motions as regards the evolution of the eccentricity $e$. Deferring details to a separate study, we here report only the main phenomenon of relevance for the re-entry mechanism, namely the fact that, at each altitude, increasing the inclination leads to a Kozai-Lidov type bifurcation (similar in nature as for the Earth satellites in the $2g$ resonance, see \cite{legnaro2023orbital}), which turns the central fixed point of the resonance from stable to unstable, generating a figure-8 separatrix along with a pair of new stable fixed points corresponding to \textit{frozen orbits} of constant eccentricity. At even higher inclinations, additional frozen orbits can also be generated by a second (pitchfork) bifurcation, which turns the central fixed point from unstable to stable. 

\subsection{Eccentricity growth and the re-entry mechanism}
\label{ssec:reentry}
Based on the phase portraits of the integrable Hamiltonian $\mathcal{H}_I$, as shown above, we can now obtain a theoretical prediction on the borders of initial conditions in the plane $(i,e)$, for fixed altitude, separating the initial conditions of orbits leading to re-entry from those which do not. This prediction is shown by the thick green dots in all three panels of Fig.\ref{fig:LT_maps}, and they are computed as follows.
\begin{itemize}
    \item For a given altitude (value of $a$), and for each fixed value of $J_F$ (or $i_0$), we first locate all the \textit{stable} fixed points within the corresponding phase portrait of $\mathcal{H}_I$ which are \textit{inside} the disc $e=e_{re}(a)$. Such can be either the central fixed point of the resonance (see top row of Fig.\ref{fig:500kmfli}, which corresponds to the value $i_0=55^\circ$ for $a-r_L=500~km$), or the stable fixed point of the frozen orbit generated after the Lidov-Kozai bifurcation (see top row of Fig.\ref{fig:500kmfli}, which corresponds to the value $i_0=56^\circ$ for $a-r_L=1500~km$)). In either case, provided that the stable fixed point is within the disc $e=e_{re}(a)$, we compute the outermost closed invariant curve $\mathcal{C}(e,g;J_F)$ around the point which comes tangent to the circumference of the disc. The area inside this curve corresponds, now, to orbits of bounded eccentricity $e<e_{re}$.  
    \item The curve $\mathcal{C}(e,g;J_F)$ intersects each of the the planes $(i,e)$ of Figure \ref{fig:LT_maps} at one point. The initial choice of the angle $\omega$ determines a `scanning direction' (e.g. $g=0$ in Figure \ref{fig:LT_maps} corresponds to a vertical scanning direction). 
    The intersection of such line with $\mathcal{C}(e,g;J_F)$ yields a unique value of the eccentricity for each value of $J_F$ and this can be transformed to a unique point in the plane $(i,e)$ solving the fast drift plane equation (Eq. (\ref{eq:planefd})).
    \item Iterating the computation for different values of the parameter $J_F$ yields the ensemble of points marking the theoretical borders of the re-entry domain.
\end{itemize}

\begin{figure}
    \centering
    \includegraphics[width = 0.9\textwidth]{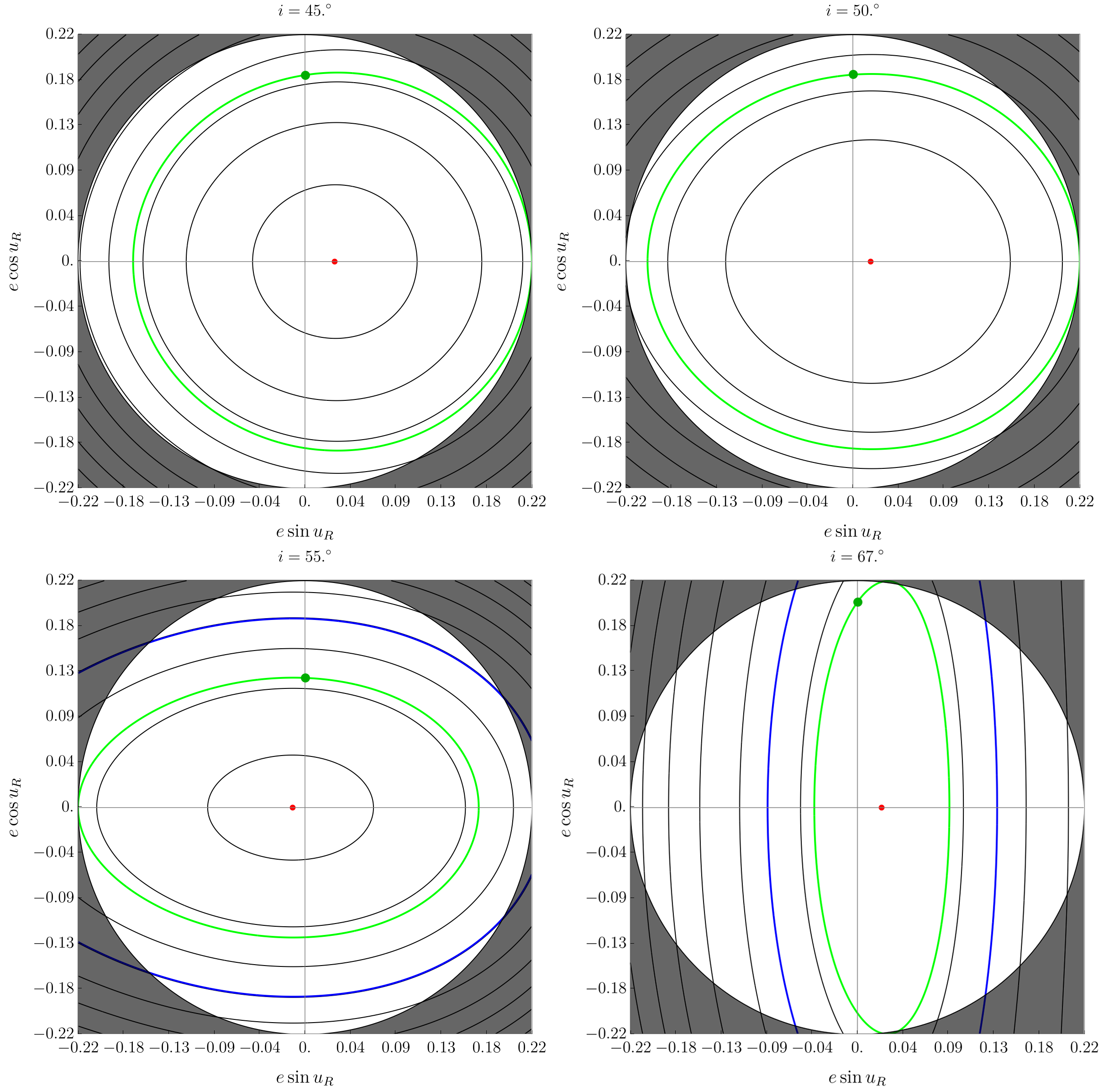}
    \caption{
    Visualization of the theoretical prediction of the borders of initial conditions in the plane $(i,e)$, for fixed altitude, separating the initial conditions of orbits leading to re-entry from those which do not detailed in \ref{ssec:reentry}. Figure  
    \ref{fig:LT_maps} shows such predictions for $\omega_0 = 0$, which corresponds to the vertical scanning direction. Thus, we find a value of the eccentricity for a given value of $J_F$ by intersecting $\mathcal{C}(e,g;J_F)$ (green curve) with the vertical axis.}
    \label{fig: analytic_predictions_visualization}
\end{figure}

\begin{figure}
    \centering
    \includegraphics[width = 0.9\textwidth]{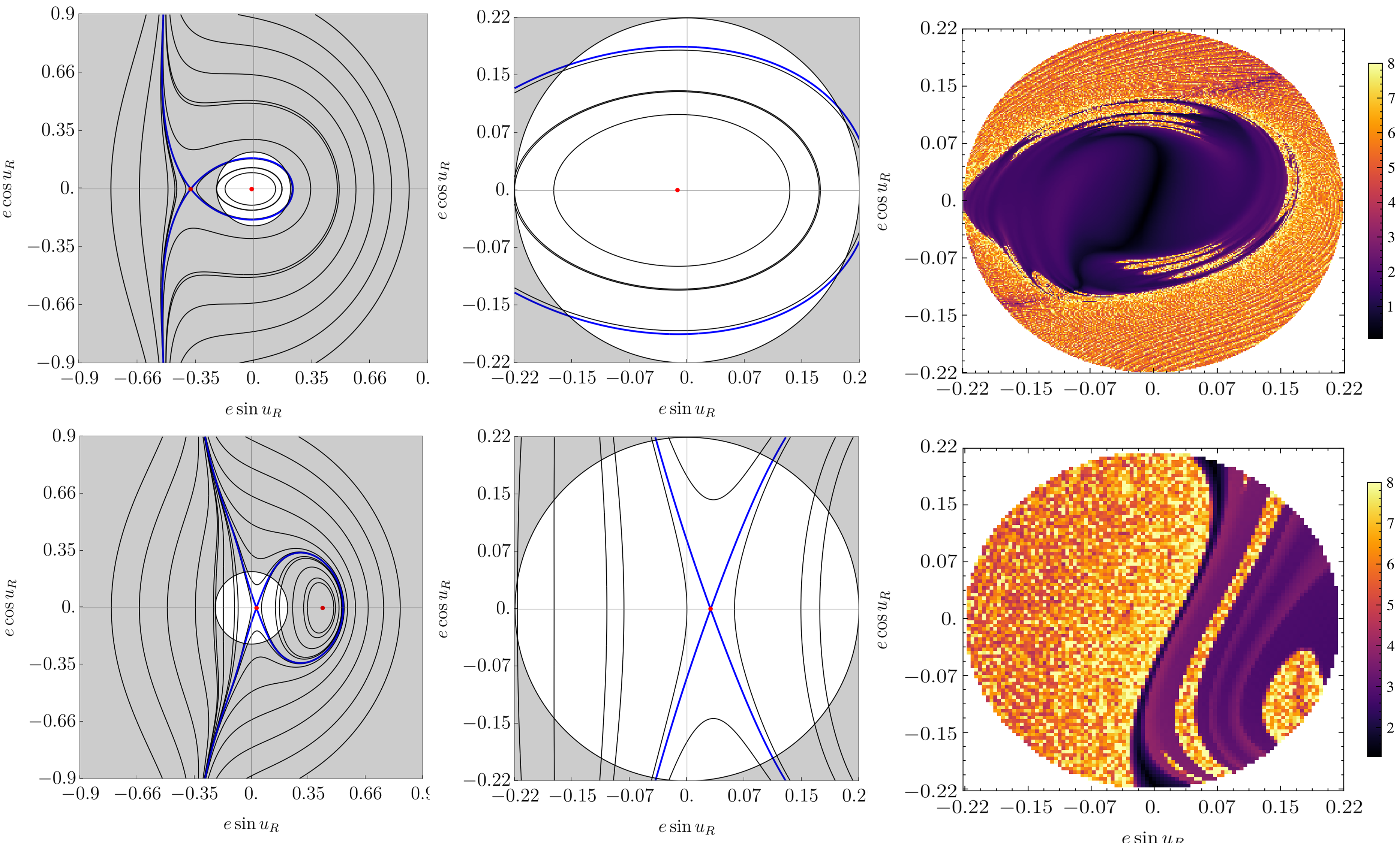}
    \caption{Two phase portraits (left)of the integrable resonance model for the altitude $a-r_L=500$~km, at the label values of the inclination $i_0=55^{\circ}$ (top) and $i_0=65^\circ$ (bottom). The middle panels show a zoom to the disc $e<e_{re}$, and the right panels the numerical FLI maps for the same initial conditions (see text). }
    \label{fig:500kmfli}
\end{figure}

A visualization of the above process is shown in Figure \ref{fig: analytic_predictions_visualization}. The closed green curves are tangent to the circles $e=e_{re}$ and mark the border of motions protected from re-entry. The thick points on these curves correspond to the points of a chosen fixed argument of perilune $\omega_0=g=-u_R$. The distance of this point from the origin $(0,0)$ yields the eccentricity of one point $(i,e)$ in the theoretical curve marking the border of re-entry in the lifetime maps. 

\begin{figure}
    \centering
    \includegraphics[width = 0.9\textwidth]{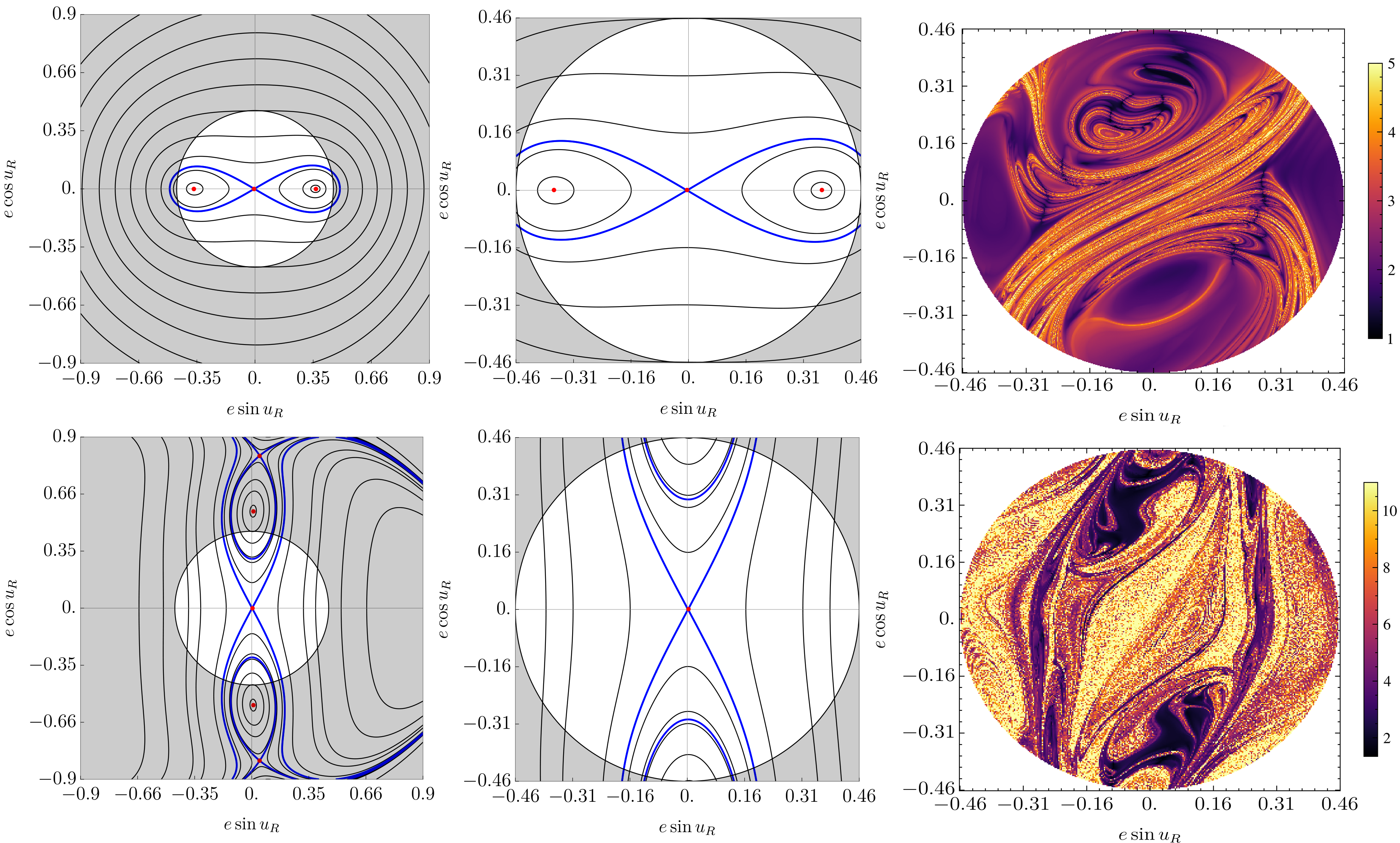}
    \caption{Same as in Fig.\ref{fig:500kmfli}, but for the altitude $a-r_L=1500~$km. The label values of the inclinations are $i_0=56^\circ$ (top) and $i_0=78^\circ$ (bottom).}
    \label{fig:1500kmfli}
\end{figure}
In all three panels of Fig.\ref{fig:LT_maps} we observe that the theoretical prediction for the borders follows closely the limits of the numerical domain of re-entry obtained with the full model $\mathcal{H}_{SM}$. This indicates that the chains of bifurcations generating new stable fixed points (and, hence, frozen orbits) as predicted by the integrable model $\mathcal{H}_I$ roughly corresponds to the sequence of emergence of frozen orbits in the real problem. Figures \ref{fig:500kmfli} and \ref{fig:1500kmfli} show that this is essentially correct. The right panels in these figures provide FLI maps for the same initial conditions as in the phase portraits of the integrable approximation. These figures show that the stability character of the fixed points in the integrable approximation is essentially reproduced by the FLI maps made with the complete force model. However, in the FLI maps we see several chaotic structures formed around the unstable fixed points, which are not present in the phase portraits of the integrable model. Similarly as exposed in \cite{daquin2021deep}, one can show that such structures represent the stable manifolds of the normally hyperbolic invariant manifold of circular orbits at the resonance. In practical terms, such manifolds render fuzzy the border separating bounded from re-entry orbits, as readily recognised in Figure \ref{fig:LT_maps}.

\section{Conclusions}
In the sections above, we discussed the main features of secular dynamics for lunar satellites, by constructing various models of secular equations of motion, all stemming from simplifications of the secular model proposed in the final report \cite{efthymiopoulos2023selena} on the semi-analytic propagator for lunar satellite orbits SELENA. In particular, we discussed the emergence of secular resonances and their role in the phenomenon of eccentricity growth and re-entry (collision with the Moon) for lunar satellites. Our main conclusions can be summarized as follows: 
\begin{itemize}
    \item As a result of the strong influence of lunar mascons, the character of the orbits is what we call `essentially non-secular' at altitudes $a-r_L<100~km$. This is defined as the altitude where the short-period variations in semi-major axis exceed a value of $1\%$. On the other hand, based on the relative importance of the multipole harmonics of the lunar potential or the tidal forces by third bodies (essentially only by the Earth), we distinguish three zones of `essentially secular dynamics': the low, middle, and high altitude zone. 
    \item A careful comparison of the contributions of all secular terms produced either by the lunar potential, or by the Earth's tidal potential, shows that the only secular resonance with well developed separatrices present in all three altitude zones is the $2g$ one. Most other resonances are supressed, due to the high value of the coefficients of the Moon's zonal harmonics compared to the small value of the Moon's obliquity angle. 
    \item We create a simplified model (Hamiltonian $\mathcal{H}_{SSM}$; see subsection \ref{ssec: hamssm}), with few harmonics (Eq.(\ref{eq:ssmset})) which accurately represents the secular dynamics of the complete model and allows to study independently the role of each potential term in it. 
    \item We further construct an integrable model, whose sequences of bifurcations of frozen orbits allow to obtain theoretically the borders separating domains of initial conditions leading to a satellite's re-entry by eccentricity growth, from those which do not. These borders were compared with numerical lifetime maps, obtained with the full equations of motion, showing a very good agreement. 
\end{itemize}  
    
\clearpage
\appendix
\section{GRAIL coefficients}
Here, we report in the table below the numerical values of the $C_{nm}$ and $S_{nm}$ coefficients of the lunar potential expansion up to order 10, and their respective normalized values $\bar{C}_{nm}$ and $\bar{S}_{nm}$ obtained by multiplying the coefficients with the factor
\begin{equation}
    \sqrt{\frac{s (1 + 2n) (n-m)!}{(n+m)!}}, \quad {\text{$s = 1$ if $n=0$, otherwise $s = 2$}}.
\end{equation}
The radius of the Moon is $r_L = 1738.0$ km, and the gravitational parameter of the Moon is $\mu_L = {\cal G}M_L=0.490280012616\times10^4 \; \si{km}^3/\si{s}^2.$
\\

\label{app: GRAIL}
\begin{longtable}{|cc|cc|cc|}
    \toprule
    $n$ & $m$ & $C_{nm}$ & $S_{nm}$ & $\bar{C}_{nm}$ & $\bar{S}_{nm}$ \\
    \midrule
    \endfirsthead
    \toprule
    $n$ & $m$ & $C_{nm}$ & $S_{nm}$ & $\bar{C}_{nm}$ & $\bar{S}_{nm}$ \\
    \midrule
    \endhead
    \midrule
    \multicolumn{6}{r}{Continued on next page} \\
    \midrule
    \endfoot
    \bottomrule
    \endlastfoot
    1 & 0 & 0.0000e+00 & 0.0000e+00 & 0.0000e+00 & 0.0000e+00 \\
    1 & 1 & 0.0000e+00 & 0.0000e+00 & 0.0000e+00 & 0.0000e+00 \\
    \midrule
    2 & 0 & -9.0884e-05 & 0.0000e+00 & -2.0322e-04 & 0.0000e+00 \\
    2 & 1 & 1.4664e-11 & 1.1733e-09 & 1.8931e-11 & 1.5147e-09 \\
    2 & 2 & 3.4673e-05 & 9.0792e-10 & 2.2381e-05 & 5.8606e-10 \\
    \midrule
    3 & 0 & -3.1973e-06 & 0.0000e+00 & -8.4593e-06 & 0.0000e+00 \\
    3 & 1 & 2.6368e-05 & 5.4545e-06 & 2.8481e-05 & 5.8916e-06 \\
    3 & 2 & 1.4172e-05 & 4.8780e-06 & 4.8405e-06 & 1.6662e-06 \\
    3 & 3 & 1.2275e-05 & -1.7742e-06 & 1.7117e-06 & -2.4740e-07 \\
    \midrule
    4 & 0 & 3.2348e-06 & 0.0000e+00 & 9.7043e-06 & 0.0000e+00 \\
    4 & 1 & -6.0135e-06 & 1.6643e-06 & -5.7049e-06 & 1.5789e-06 \\
    4 & 2 & -7.1162e-06 & -6.7770e-06 & -1.5912e-06 & -1.5154e-06 \\
    4 & 3 & -1.3499e-06 & -1.3445e-05 & -8.0670e-08 & -8.0349e-07 \\
    4 & 4 & -6.0070e-06 & 3.9264e-06 & -1.2692e-07 & 8.2961e-08 \\
    \midrule
    5 & 0 & -2.2379e-07 & 0.0000e+00 & -7.4221e-07 & 0.0000e+00 \\
    5 & 1 & -1.0116e-06 & -4.1189e-06 & -8.6628e-07 & -3.5272e-06 \\
    5 & 2 & 4.3995e-06 & 1.0571e-06 & 7.1200e-07 & 1.7108e-07 \\
    5 & 3 & 4.6614e-07 & 8.6989e-06 & 1.5399e-08 & 2.8736e-07 \\
    5 & 4 & 2.7543e-06 & 6.7678e-08 & 2.1445e-08 & 5.2696e-10 \\
    5 & 5 & 3.1106e-06 & -2.7544e-06 & 7.6589e-09 & -6.7820e-09 \\
    \midrule
    6 & 0 & 3.8184e-06 & 0.0000e+00 & 1.3768e-05 & 0.0000e+00 \\
    6 & 1 & 1.5283e-06 & -2.5996e-06 & 1.2024e-06 & -2.0454e-06 \\
    6 & 2 & -4.3973e-06 & -2.1677e-06 & -5.4704e-07 & -2.6967e-07 \\
    6 & 3 & -3.3175e-06 & -3.4274e-06 & -6.8785e-08 & -7.1064e-08 \\
    6 & 4 & 3.4124e-07 & -4.0581e-06 & 1.2917e-09 & -1.5362e-08 \\
    6 & 5 & 1.4544e-06 & -1.0342e-05 & 1.1738e-09 & -8.3465e-09 \\
    6 & 6 & -4.6842e-06 & 7.2299e-06 & -1.0913e-09 & 1.6844e-09 \\
    \midrule
    7 & 0 & 5.5934e-06 & 0.0000e+00 & 2.1663e-05 & 0.0000e+00 \\
    7 & 1 & 7.4717e-06 & -1.1973e-07 & 5.4687e-06 & -8.7635e-08 \\
    7 & 2 & -6.5015e-07 & 2.4111e-06 & -6.4756e-08 & 2.4015e-07 \\
    7 & 3 & 5.9942e-07 & 2.3573e-06 & 8.4434e-09 & 3.3205e-08 \\
    7 & 4 & -8.4367e-07 & 7.5653e-07 & -1.7916e-09 & 1.6065e-09 \\
    7 & 5 & -2.0686e-07 & 1.0693e-06 & -7.3213e-11 & 3.7845e-10 \\
    7 & 6 & -1.0652e-06 & 1.1005e-06 & -7.3938e-11 & 7.6385e-11 \\
    7 & 7 & -1.8204e-06 & -1.6003e-06 & -3.3770e-11 & -2.9686e-11 \\
    \midrule
    8 & 0 & 2.3468e-06 & 0.0000e+00 & 9.6762e-06 & 0.0000e+00 \\
    8 & 1 & 4.1684e-09 & 1.0980e-06 & 2.8644e-09 & 7.5455e-07 \\
    8 & 2 & 3.0093e-06 & 1.9306e-06 & 2.4717e-07 & 1.5857e-07 \\
    8 & 3 & -1.8890e-06 & 9.5448e-07 & -1.9098e-08 & 9.6498e-09 \\
    8 & 4 & 3.4087e-06 & -5.2825e-07 & 4.4490e-09 & -6.8947e-10 \\
    8 & 5 & -1.2481e-06 & 2.9186e-06 & -2.2590e-10 & 5.2826e-10 \\
    8 & 6 & -1.6604e-06 & -2.1147e-06 & -4.6373e-11 & -5.9060e-11 \\
    8 & 7 & -1.5097e-06 & 3.2689e-06 & -7.6980e-12 & 1.6668e-11 \\
    8 & 8 & -2.4857e-06 & 2.1163e-06 & -3.1687e-12 & 2.6978e-12 \\
    \midrule
    9 & 0 & -3.5309e-06 & 0.0000e+00 & -1.5391e-05 & 0.0000e+00 \\
    9 & 1 & 1.8670e-06 & 8.1043e-08 & 1.2131e-06 & 5.2660e-08 \\
    9 & 2 & 1.9278e-06 & -1.3876e-06 & 1.3353e-07 & -9.6113e-08 \\
    9 & 3 & -1.9924e-06 & 2.2018e-06 & -1.5058e-08 & 1.6640e-08 \\
    9 & 4 & -1.8844e-06 & -1.4258e-06 & -1.6126e-09 & -1.2201e-09 \\
    9 & 5 & -1.5625e-06 & -3.5247e-06 & -1.5982e-10 & -3.6051e-10 \\
    9 & 6 & -2.1272e-06 & -3.0026e-06 & -2.8088e-11 & -3.9648e-11 \\
    9 & 7 & -3.9148e-06 & -1.0688e-07 & -7.4612e-12 & -2.0371e-13 \\
    9 & 8 & -1.3120e-06 & -2.2035e-06 & -4.2883e-13 & -7.2021e-13 \\
    9 & 9 & -9.3820e-07 & 2.4881e-06 & -7.2280e-14 & 1.9169e-13 \\
    \midrule
    10 & 0 & -1.0693e-06 & 0.0000e+00 & -4.9001e-06 & 0.0000e+00 \\
    10 & 1 & 8.4160e-07 & -9.5407e-07 & 5.2004e-07 & -5.8953e-07 \\
    10 & 2 & 3.5724e-07 & -2.6511e-07 & 2.1241e-08 & -1.5763e-08 \\
    10 & 3 & 4.8420e-07 & 6.6884e-07 & 2.8231e-09 & 3.8996e-09 \\
    10 & 4 & -3.5730e-06 & 1.5789e-06 & -2.1043e-09 & 9.2992e-10 \\
    10 & 5 & 6.9970e-07 & -3.1458e-07 & 4.3439e-11 & -1.9530e-11 \\
    10 & 6 & -1.2729e-07 & -2.0953e-06 & -8.8353e-13 & -1.4543e-11 \\
    10 & 7 & -3.9986e-06 & -9.1075e-07 & -3.3657e-12 & -7.6659e-13 \\
    10 & 8 & -3.5594e-06 & 2.8489e-06 & -4.0771e-13 & 3.2632e-13 \\
    10 & 9 & -4.7532e-06 & -5.1547e-08 & -8.8320e-14 & -9.5782e-16 \\
    10 & 10 & 9.4793e-07 & -1.7194e-06 & 3.9386e-15 & -7.1439e-15 \\
    \end{longtable}

\clearpage
\bibliographystyle{elsarticle-num}

\end{document}